\documentclass[12pt]{iopart}
\usepackage{graphicx}
\usepackage{amssymb}
\usepackage{fontenc}
\usepackage{float}

\usepackage{color}
\definecolor{correction}{rgb}{0,0,0} 
\definecolor{correction2}{rgb}{0,0,0}
\newcommand{\cor}[1]{\textcolor{correction2}{#1}}
\begin{document}

\newcommand{\ud}{\mathrm{d}}
\eqnobysec
\title{Physics of plasma burn-through and DYON simulations for the \textcolor{correction}{JET} ITER-like wall}
\author{Hyun-Tae Kim$^{1,2}$, A.C.C. Sips$^{1,3}$, and EFDA-JET contributors*}
\address{JET-EFDA Culham Science Centre, Abingdon, OX14 3DB, UK.}
\address{$^1$Department of Physics, Imperial College London, Prince Consort Road, London, SW7~2AZ, UK}
\address{$^2$EURATOM/CCFE Fusion Association, Abingdon, OX14~3DB, UK}
\address{$^3$JET/EFDA, Culham Science Centre, Abingdon, OX14~3DB, UK}
\address{*See the Appendix of F. Romanelli et al., Proceedings of the 24th IAEA Fusion Energy Conference 2012, San Diego, US}
\ead{hyun.kim09@imperial.ac.uk}


\begin{abstract}
This paper presents the DYON simulations of the plasma burn-through phase at Joint European Torus (JET) with the ITER-like wall. The main purpose of the study is to \textcolor{correction}{validate} the simulations with the ITER-like wall, made of beryllium. Without impurities, the burn-through process of a pure deuterium plasma is described using DYON simulations, and the criterion for deuterium burn-through is derived analytically. The plasma burn-through with impurities are simulated using \textcolor{correction}{wall-sputtering} models in the DYON code, which are modified for the ITER-like wall.  The \textcolor{correction}{wall-sputtering} models and the validation against JET data are presented. The impact of the assumed plasma parameters in DYON simulations are discussed by means of parameter scans. As a  result, the operation space \textcolor{correction}{of prefill gas pressure and toroidal electric field} for plasma burn-through in JET is compared to the Townsend avalanche criterion.
\end{abstract}

\maketitle


\section{Introduction}
\subsection{Motivation}  
Tokamak start-up consists of the electron avalanche phase, the plasma burn-through phase, and the ramp-up phase of the plasma current $I_p$ \cite{ITERphysics1999}. The Townsend avalanche theory\cite{tanga}\cite{lloyd_D} is generally used to calculate the required electric field for electron avalanche ($E_{avalanche}$) at a given prefill gas pressure $p$ and effective connection length $L_f$ as shown below,
\begin{eqnarray}
E_{avalanche}\cor{[V/m]} \geq \frac{1.25 \times 10^4  \times  p \cor{[Torr]}}{\ln(510 \times p\cor{[Torr]} \times L_{f}\cor{[m]})}, \label{TownsendCriterion}
\end{eqnarray}
The required $E_{avalanche}$ for plasma initiation in the International Thermonuclear Experimental Reactor (ITER) has also been calculated using \Eref{TownsendCriterion} as presented in the ITER Physics Basis\cite{ITERphysics1999}. 

However, the Townsend avalanche theory is not sufficient to explain all non-sustained break-down discharges. In order for plasma current to increase, sufficient ionization of the prefill gas (deuterium) and impurities, i.e. plasma burn-through, is necessary. Otherwise, most heating power is lost through radiation and ionizations of the remaining neutrals, so that it prevents electron temperature from increasing in the ramp-up phase of plasma current\cite{tanga}. It should be noted that the required loop voltage for plasma burn-through, the burn-through criterion, is generally higher than that for electron avalanche in present tokamaks\cite{ITERphysics1999}, and a number of start-up failures in current devices result from the failure of plasma burn-through. For  experiments with the carbon wall in Joint European Torus (JET), more than $85\%$ of all non-sustained breakdown failures occurred during the plasma burn-through phase \cite{Peter_2012IAEA}. These start-up failures could be reduced by understanding key physics aspects of the plasma burn-through phase.

Furthermore, due to the engineering issues in ITER, resulting from the use of superconducting central solenoid coils and a continuous vacuum vessel, the maximum toroidal electric field on-axis is limited up to 0.35 $[V/m]$ \cite{ITERphysics2007}, which is much lower than the typical toroidal electric field used for plasma burn-through in current devices, e.g. $\sim$ $1$ $[V/m]$ in JET. Tokamak start-up using such a low electric field limits the operation space available for the range of prefill gas pressure, magnetic error fields, and impurity content\cite{lloyd_ITER}. In order to obtain more confidence in the start-up scenario at ITER,  a predictive simulation for plasma burn-through is required. 

For reliable start-up using a low electric field, RF-assisted start-up using Electron Cyclotron Heating (ECH)\cite{JStober} or Ion Cyclotron Heating (ICH)\cite{Koch1} is planned in ITER. However, launching excessive RF power into the vacuum vessel without a plasma or with a very low temperature plasma can result in serious \cor{damage} to the diagnostic systems, due to the low absorption efficiency of the RF power. Hence, in order to estimate the required ECH power (although not presented in this paper)  understanding the plasma burn-through conditions (or requirements) is also important.

The DYON \textcolor{correction}{(DYnamic 0D model of Non-fully ionized plasma)} code is a plasma burn-through simulator, developed at JET. In the DYON code, the confinement time is modelled as a function of plasma parameters considering parallel transport as well as perpendicular transport, and the impurity influx is calculated  using \textcolor{correction}{wall-sputtering} models. \textcolor{correction}{The impurity densities in each charge state are calculated self-consistently by particle balances in non-coronal equilibrium. In this calculation, neutral screening effects are also taken into account, and the atomic reactions and the resultant radiation are computed using Atomic Data and Analysis Structure (ADAS) package\cite{Summers}.} A detailed description on DYON code can be found in \cite{DYON}. 

The DYON simulation results show a reasonable agreement with JET data with the carbon wall\cite{DYON}. The recent installation of ITER-Like Wall at JET i.e. a combination of beryllium and tungsten protection tiles \cite{ILW_VPhilipps}, enabled us to validate the \textcolor{correction}{wall-sputtering} models for a beryllium wall with results from recent experiments. In this article, the \textcolor{correction}{modified wall-sputtering model} in the DYON code will be explained, and the simulation results are compared to JET data with the ITER-like wall. \textcolor{correction}{Using the validated models, the operation window of prefill gas pressure and toroidal electric field for plasma burn-through in JET is computed, and it is compared with the Townsend criterion.}   

\subsection{Structure of the paper}
Plasma burn-through of a pure deuterium plasma will be discussed in section \ref{Physicsofdeuteriumburn-through}.  In section \ref{ReviewofPSImodelsforcarbonwall}, the previous \textcolor{correction}{wall-sputtering} model for the carbon wall is reviewed.  In section \ref{NewPSImodelsforITER-Likewall}, the modification of the \textcolor{correction}{wall-sputtering}  model for the JET ITER-like wall will be introduced. In section \ref{ILWsimulationsandcomparisontotheJETdata}, DYON simulation results for the JET ITER-like wall are compared to JET data to provide a validation. The effects of the paramters assumed in the DYON simulation, i.e. deuterium recycling coefficient, fuelling efficiency, and initial carbon content, will be investigated in section \ref{ILWsimulationsandcomparisontotheJETdata}.  In section \ref{Operationspaceforplasmaburn-through}, the computed criterion for plasma burn-through with impurities is compared to the Townsend avalanche criterion, and the operation space for JET is presented. Discussion and conclusions are given in section \ref{sectionDiscussion} and section \ref{sectionConclusion}, respectively.

\section{Physics of deuterium burn-through}\label{Physicsofdeuteriumburn-through}

\textcolor{correction}{Impurities during the plasma burn-through phase result from complex plasma surface interaction. This makes analytical investigation on plasma burn-through extremely complicated. To gain an insight into the key physics aspects in the plasma burn-through phase, it is worth starting the investigation of a pure deuterium plasma. Furthermore, according to recent observation in JET with the ITER-like wall, deuterium radiation can be critical for plasma burn-through with beryllium wall. }

\begin{figure}[H]
\begin{center}
\includegraphics[width=1\textwidth]{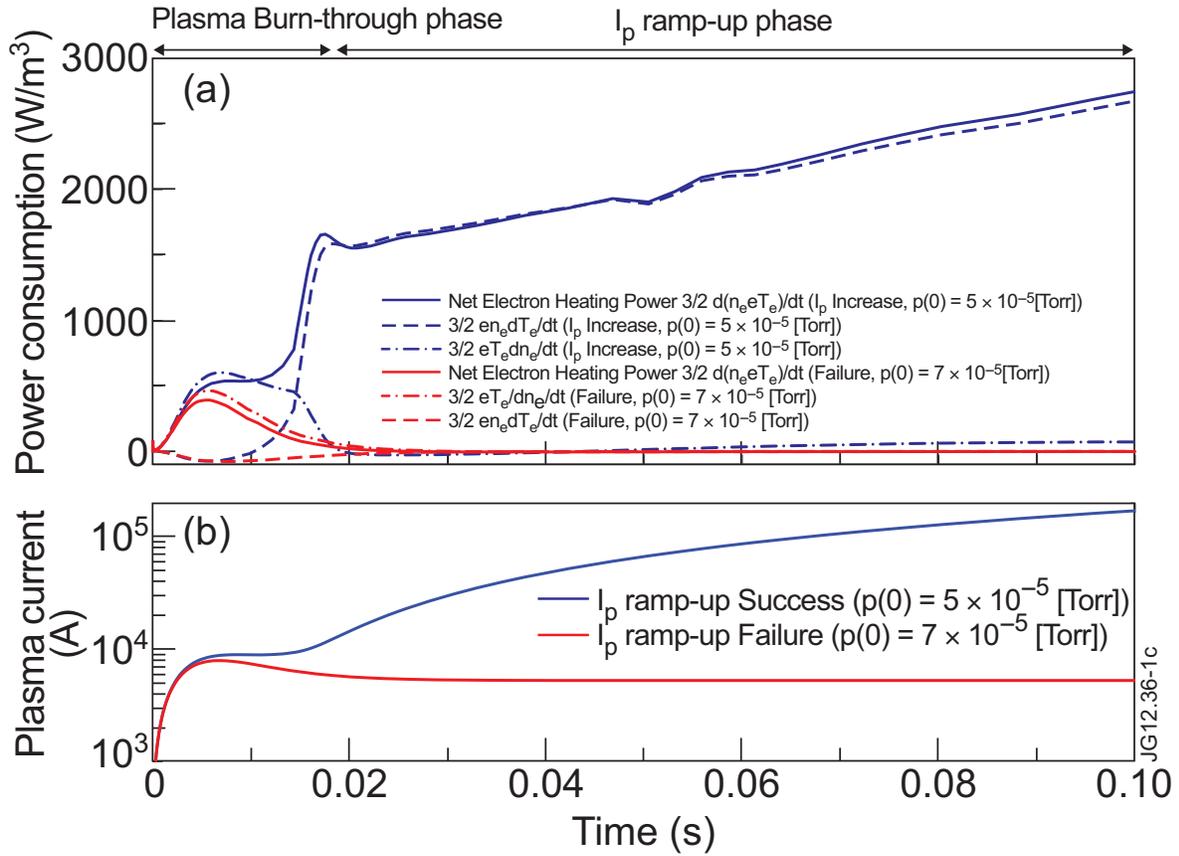} 
\caption{DYON simulation results for a pure deuterium plasma. The colors of lines in (a) and (b) indicate successful $I_p$ ramp-up(blue) and failure(red). The solid lines represent the net electron heating power $P_e$. The dashed lines and the chain lines are the amount of $P_e$ consumed by increasing $T_e$ and increasing $n_e$, respectively. The corresponding plasma currents $I_p$ are represented by the blue solid line($I_p$ ramp-up) and the red solid line(non-sustained break-down) in (b). In order for $I_p$ to increase, $P_e$ must be positive in the $I_p$ ramp-up phase.} \label{Figure1}
\end{center}
\end{figure}

The results presented in this section are obtained by DYON simulations  (Figure \ref{Figure1}, \ref{Figure2},  and \ref{Figure4}) without any impurity model. Figure \ref{Figure3_1} is drawn using an analytical formula. The plasma parameters assumed in the DYON simulations of a pure deuterium plasma are given in Table \ref{defaultvalues}.  In order to simulate the cases of successful and failed plasma burn-through, two different prefill gas pressures are assumed, $5 \times 10^{-5}[Torr]$(Success), and $7 \times 10^{-5}[Torr]$(Failure)  in Figures \ref{Figure1} and \ref{Figure2}.\footnote{1[Torr] = 1.33322368 [mbar]} In Figure \ref{Figure4}, a wider range of prefill gas pressures ($1 \times 10^{-5}$, $3 \times 10^{-5}$, $5 \times 10^{-5}$, and $7 \times 10^{-5}[Torr]$) is used to show the effects of prefill gas pressure on plasma burn-through.

\subsection{Condition for plasma current ramp-up}

Assuming no eddy current in the passive structure, the plasma current $I_p$ in tokamaks can be calculated with the circuit equation
\begin{eqnarray}
I_p=\frac{1}{R_p}(U_l-L_p\frac{dI_p}{dt})
\end{eqnarray}
where $R_p$, $U_l$, and $L_p$ are plasma resistance, loop voltage, and plasma inductance, respectively. In order for $I_p$ to increase for a given $U_l$, which is approximately constant in the $I_p$ ramp-up phase, $R_p$ must be decreasing continuously. According to Spitzer resistivity,  $R_p$ decreases as $T_e$ increases \cite{wesson}, i.e.  $R_p \propto T^{-\frac{3}{2}}_e$. Therefore, 
\begin{eqnarray}
\frac{dT_e}{dt} > 0 
\end{eqnarray}
is a necessary condition for $I_p$ ramp-up. 

Whether or not $T_e$ increases is determined by the equation for electron energy balance,
\begin{eqnarray}
P_e=\frac{3}{2}\frac{d(n_e k T_e)}{dt} = \frac{3}{2} k T_e\frac{dn_e}{dt} + \frac{3}{2} n_e \frac{dkT_e}{dt}, \label{consumptionofPe}
\end{eqnarray}
where $P_e$ is the net electron heating power, determined by the ohmic heating power $P_{Oh}$ (for cases without assist by additional heating) and the total electron power loss $P_{Loss}$, i.e. $P_e=P_{Oh}-P_{Loss}$. As separated into the two terms in \Eref{consumptionofPe}, the net electron heating power $P_e$ is consumed by increasing $n_e$ or $T_e$, i.e. $\frac{3}{2} k T_e\frac{dn_e}{dt}$ or $\frac{3}{2} n_e \frac{dkT_e}{dt}$.

The change of \Eref{consumptionofPe} during the plasma burn-through phase is described in Figure \ref{Figure1} using the DYON simulation results: (a) the power consumption for successful $I_p$ ramp-up (blue) and failed case (red), and (b) the corresponding plasma current in each case. As shown, $P_e$ is positive for the successful case, and goes to zero in the failed case during the $I_p$ ramp-up phase. Whereas the power consumed by the increasing $n_e$ (chain lines) is dominant in the plasma burn-through phase, it is small enough to be ignored in the $I_p$ ramp-up phase as shown in Figure \ref{Figure1}(a), \textcolor{correction}{i.e. blue solid line $\approx$ blue dashed line.} Therefore, $P_e$ in the $I_p$ ramp-up phase can be approximated to be $P_e \approx \frac{3}{2}n_e\frac{dkT_e}{dt}$. Accordingly, in order for $T_e$ increases, $P_e$ must be positive in the $I_p$ ramp-up phase, i.e.
\begin{eqnarray}
P_e > 0. \label{Pe>0}
\end{eqnarray}
In this simulation, the deuterium recycling coefficient $Y^D_D$ is assumed as $1$. In the case that $Y^D_D$ is higher than $1$, the power consumed by the increasing $n_e$ would not be $0$. However, \Eref{Pe>0} is still a necessary condition for the increase in $T_e$ unless the power consumed by the increasing $n_e$ becomes significant.

\begin{figure}[H]
\begin{center}
\includegraphics[width=1\textwidth]{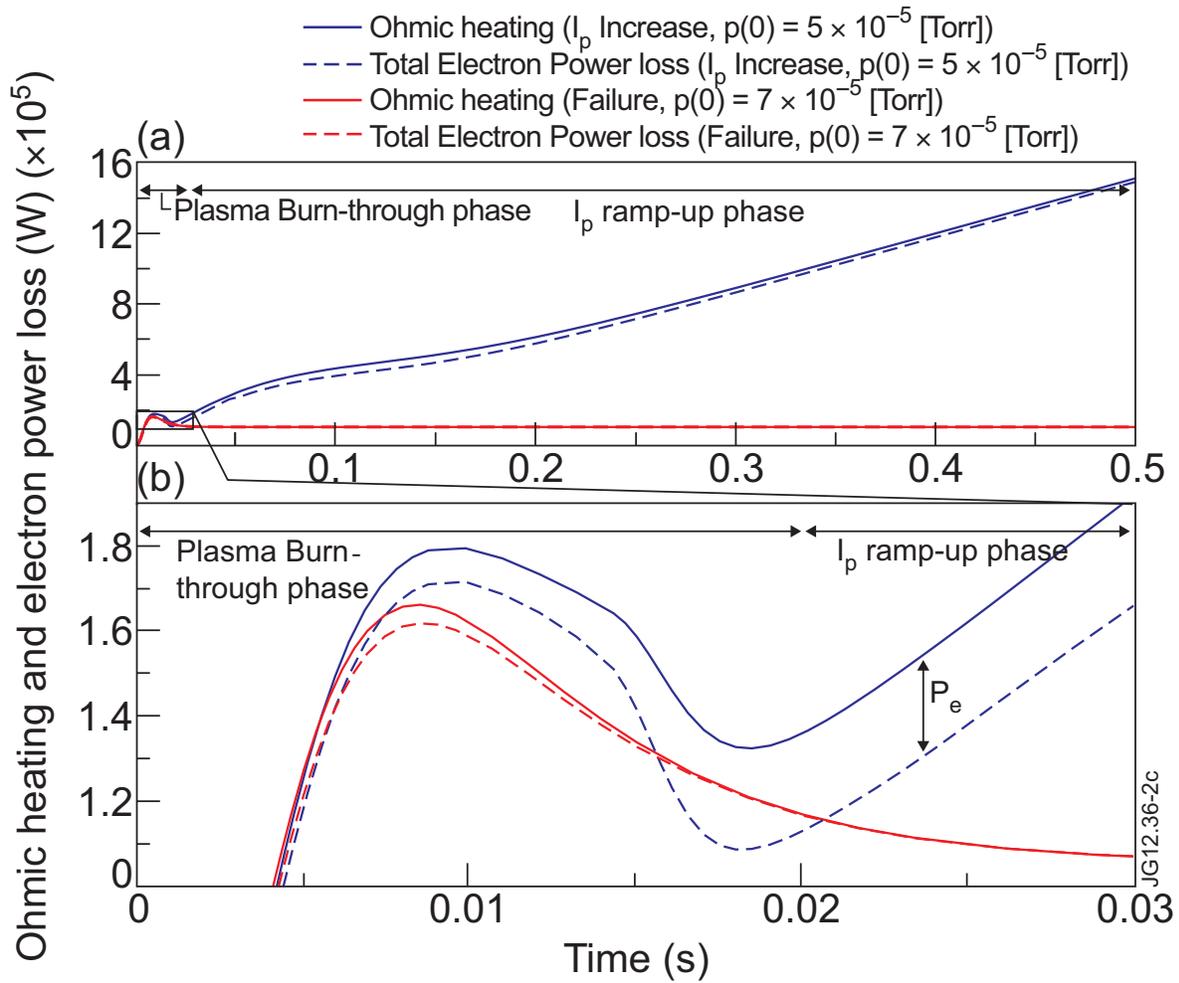}
\caption{DYON simulation results for a pure deuterium plasma. The solid lines and the dashed lines in (a) show $P_{Oh}$  and  $P_{Loss}$ in successful(blue) and failure(red) cases, respectively. (b) is an expanded figure from the burn-through phase in (a). It is determined by $P_e$ during the plasma burn-through phase whether $P_e$ is positive for the $I_p$ ramp-up phase.} \label{Figure2}
\end{center}
\end{figure}

Figure \ref{Figure2} shows DYON simulation results for $P_{Oh}$ and $P_{Loss}$ in the case of $I_p$ ramp-up success(blue) and failure(red), respectively. In the successful case, $P_{Oh}$(blue solid line) exceeds $P_{Loss}$(blue dashed line), i.e. positive $P_e$ in the $I_p$ ramp-up phase. However, $P_{Oh}$(red solid line) and $P_{Loss}$(red dashed line) overlap in the failed case, hence $P_e$ is zero. Figure \ref{Figure2}(b), which is enlarged from Figure \ref{Figure2}(a), shows that the behaviour of $P_e$ in the burn-through phase are clearly different in the two cases. It is determined by the behaviour of $P_e$ during the plasma burn-through phase whether $P_e$ in the $I_p$ ramp-up phase is positive. Hence, $P_e$ during the plasma burn-through phase should be investigated to derive the requirement of $I_p$ ramp-up, i.e. the criterion of plasma burn-through.

\subsection{Criterion for deuterium burn-through}\label{Criterionforplasmaburn-through}

The total electron power loss, $P_{Loss}$ consists of the three power loss terms, i.e. radiation and ionization power loss $P_{rad+iz}$, equilibration power loss $P_{equi}$, and convective transport power loss $P^e_{conv}$. In the case of a pure deuterium plasma assumed in this section, they are calculated as shown below\cite{lloyd_ITER}\cite{DYON}.
\begin{eqnarray}
P_{Loss} = P_{rad+iz} + P_{equi} + P^e_{conv} \nonumber \\
P_{rad+iz} = V_p \times \mathcal{P}_{RI}(T_e) n_e n_D^0 \label{RIBpowerloss} \\
P_{equi}=V_p \times 7.75 \times 10^{-34}(T_e - T_i)\frac{n_e n^{1+}_D \ln \Lambda}{T^{3/2}_e M_D}, \label{EquilibrationPower} \\
P^e_{conv} = V_p \times \frac{3}{2}\frac{n_e k T_e}{\tau_e}, \label{ElectronConvTranPowerLoss}
\end{eqnarray}
where $n_D^0$ is a deuterium atom density, $n_D^{1+}$ is an deuterium ion density, $V_p$ is a plasma volume, $M_D$ is a deuterium ion mass in $[amu]$, $\tau_e$ is the electron particle confinement time, and $\mathcal{P}_{RI}(T_e)$ is the power  loss coefficient due to the radiation and ionization, obtained from ADAS package\cite{Summers}.

In contrast to $P_{equi}$ and $P^e_{conv}$, which simply increase with $n_e$, $P_{rad+iz}$ has a maximum value at a certain degree of ionization since $n_D^0$ decreases as $n_e$ increases. In this paper, we define the peak value of $P_{rad+iz}$ as the \emph{Radiation and Ionization Barrier} (RIB), and the degree of ionization at the RIB is defined to be the \emph{Critical Degree of Ionization} for plasma burn-through, $\gamma_{iz}(t_{RIB})$. 

The RIB is of crucial importance since the required $P_{Oh}$ for $I_p$ ramp-up is mainly determined by the RIB. As will shown, the magnitude of $P_{rad+iz}$ is dominant in $P_{Loss}$ during the plasma burn-through phase. This implies that $P_{Loss}$ also has the maximum value at $\gamma_{iz}(t_{RIB})$. Hence, once the $P_{Oh}$ exceeds the $P_{Loss}$ maximum, $P_{Loss}$ decreases significantly as ionizations proceed. This enables  $T_e$ to increase, so that ionizations continue to increase up to $100\%$, i.e. full ionization. 

\begin{figure}[H]
\begin{center}
\includegraphics[width=1\textwidth]{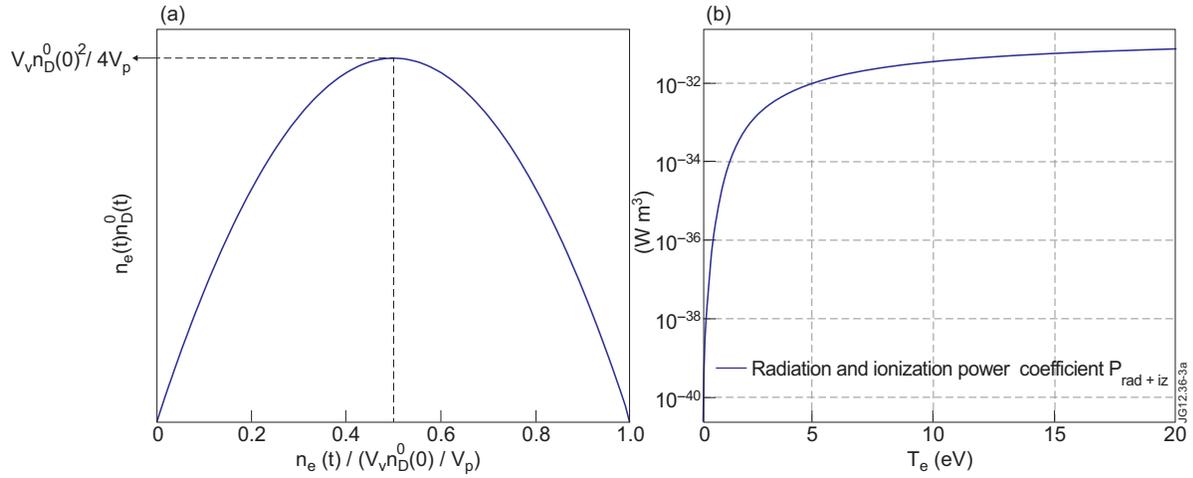}
\caption{(a) shows the change of $n_e(t) n^0_D(t)$ with the normalized $n_e$. Since $n_e(t) n^0_D(t)$ can be substituted by $n_e(t)(n^0_D(0)-n_e(t))$ in the case of a recycling coefficient($=1.0$), it has a maximum value as $n_e(t)$ approaches \Eref{maximumne}. (b) indicates electron power loss coefficient due to the radiation and ionization of deuterium, $\mathcal{P}_{RI}$, obtained from ADAS. $\mathcal{P}_{RI}$ is strongly dependent  on $T_e$ only.} \label{Figure3_1}
\end{center}
\end{figure}

During the plasma burn-through phase, the density of deuterium atoms $n^0_D(t)$ decreases, thereby increasing $n_e(t)$. When the deuterium atom density within a plasma volume decreases, neutral particles flow into the plasma volume from the ex-plasma volume, giving a dynamic fuelling effect. This effect maintains a neutral density within a plasma volume as much as the ratio of plasma volume to total neutral volume(= \textcolor{correction}{Effective vessel volume $V_V$, in which all neutrals are accessible to the plasma}). The effective reduction of neutral density in $V_p$ is $\frac{V_p}{V_V}n_e(t)$. Hence, in the case that the deuterium recycling coefficient $Y^D_D$ is $1$ and there is no gas pumping or puffing, $n^0_D$ in \Eref{RIBpowerloss} is  
\begin{eqnarray}
n^0_D(t)= n^0_{D}(0)-\frac{V_p}{V_V}n_e(t). \label{ndtbyne}
\end{eqnarray}
where $n^0_D(0)$ indicates the initial density of deuterium atoms, which is proportional to the prefill gas pressure. By substituting $n^0_D$ in \Eref{RIBpowerloss} with $n^0_D(t)$ in \Eref{ndtbyne}, $P_{rad+iz}(t)$ can be written as a \cor{quadratic} function of $n_e(t)$,
\begin{eqnarray}
P_{rad+iz}(t) = V_p \mathcal{P}_{RI}(T_e) n_e(t) n^0_D(t)  \nonumber \\
=V_p \mathcal{P}_{RI}(T_e) n_e(t) \Big( n^0_{D}(0)-\frac{V_p}{V_V}n_e(t) \Big)  \nonumber \\
=V_p \mathcal{P}_{RI}(T_e) \Big( \frac{V_V n_D^0(0)^2}{4V_p} - \frac{V_p}{V_V}\Big(n_e(t)-\frac{V_V n_D^0(0)}{2V_p}\Big)^2 \Big)  \label{Pizrad}
\end{eqnarray}
 Therefore, as ionizations proceed, $n_e(t)n^0_D(t)$ in \Eref{Pizrad} has a maximum value. Figure \ref{Figure3_1}(a) indicates the change of $n_e(t)n^0_D(t)$ with the normalized $n_e(t)$, i.e. $\frac{n_e(t)V_p}{n^0_D(0)V_V} $. As shown in Figure \ref{Figure3_1}(a), $n_e(t)n^0_D(t)$ has the maximum value,
\begin{eqnarray} 
\frac{V_V n_D^0(0)^2}{4V_p} \label{maximumndne}
\end{eqnarray}
when $n_e(t)$ is equal to 
\begin{eqnarray}
\frac{V_V n_D^0(0)}{2V_p}. \label{maximumne}
\end{eqnarray}
The power coefficient $\mathcal{P}_{RI}(T_e)$ is a function of $T_e$. Figure \ref{Figure3_1}(b) shows $\mathcal{P}_{RI}(T_e)$ obtained from ADAS package\cite{Summers}. The product of $n_e(t)n^0_D(t)$ and $\mathcal{P}_{RI}(T_e)$ results in the change of $P_{rad+iz}$, thereby the change of $P_{Loss}$ in Figure \ref{Figure2}. 

The degree of ionization in the plasma burn-through phase can be calculated with
\begin{eqnarray}
\gamma_{iz}(t)=\frac{n_e(t)}{n_e(t) + n^0_D(t)}. 
\end{eqnarray} 
The degree of ionization at the RIB, $\gamma_{iz}(t_{RIB})$, is then obtained by substituting $n_D^0(t)$ and $n_e(t)$ with \Eref{ndtbyne} and (\ref{maximumne}) as shown below.
\begin{eqnarray}
\gamma_{iz}(t_{RIB})= \frac{\frac{V_vn_d^0(0)}{2V_p}}{\frac{V_Vn_d^0(0)}{2V_p}+\Big(n_D^0(0)-\frac{V_p\frac{V_vn_d^0(0)}{2V_p}}{V_V} \Big)} \nonumber \\
= \frac{V_V}{V_V+V_p} \label{criticalDiz}
\end{eqnarray}
The plasma volume is limited by the vessel volume, i.e. $V_p \leq V_V$. This implies that $\gamma_{iz}(t_{RIB})$ is always higher than $50\%$. In the case of JET, where $V_V$ is $\sim 100[m^{3}]$ \textcolor{correction}{(Further explanation on $V_V$ is provided in the disscusion section.)} and initial plasma volume $V_p=14.8 \sim 59.2[m^{3}]$ (major radius $R=3[m]$ and minor radius $a=0.5 \sim 1[m]$), the critical degree of ionzation $\gamma_{iz}(t_{RIB})$ is $87.1 \sim 62.8\%$, respectively. 

\begin{figure}[H]
\begin{center}
\includegraphics[width=1\textwidth]{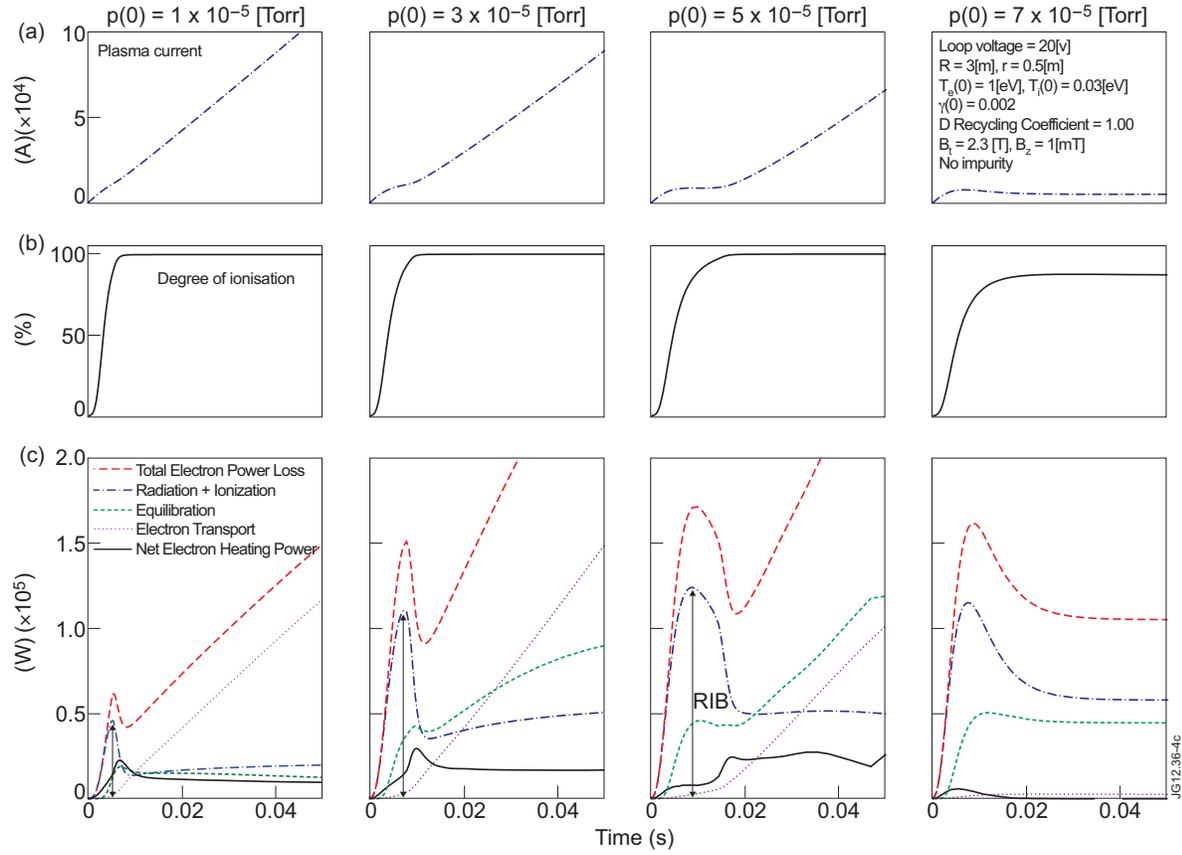}
\caption{DYON simulation results for a pure deuterium plasma. The figures show (a)the plasma current, (b)the degree of ionization, and (c)various electron power losses at different prefill gas pressures, $1 \times 10^{-5}$, $3 \times 10^{-5}$, $5 \times 10^{-5}$, and $7 \times 10^{-5}$[Torr]. The assumed loop voltage and plasma parameters are shown in Table \ref{defaultvalues}. Under the given condition, a critical prefill gas pressure for $I_p$ ramp-up exists between $5 \times 10^{-5}$ and $7 \times 10^{-5}$[Torr]. Prefill gases are almost fully ionized in the cases of successful $I_p$ ramp-up  while they are not fully ionized in the cases of failure. The colored  lines in (c) indicate $P_{Loss}$(dashed red ), $P_{equi}$(dashed green), $P^e_{conv}$(dotted cyan), and $P_{rad+iz}$(chain blue), respectively. As shown in (c), $P_{rad+iz}$ is dominant in $P_{Loss}$ during the burn-through phase, and its peak values coincide the RIB. The RIB increases with prefill gas pressure, thereby increasing $P_{Loss} maximum$. That is, the higher the prefill gas pressure, the larger the $P_{Loss}$ maximum. } \label{Figure4}
\end{center}
\end{figure}

Figure \ref{Figure4} shows DYON simulation results of (a) plasma current, (b) degree of ionization, and (c) electron power losses for different prefill gas pressures. The $I_p$ ramp-up is delayed until almost $100\%$ ionization is achieved in the low prefill gas pressure cases ($1 \times 10^{-5}$, $3 \times 10^{-5}$, and $5 \times 10^{-5}[Torr]$), and the delay is extended with increasing prefill gas pressures.  The $I_p$ ramp-up fails at a prefill gas pressure over $7 \times 10^{-5}[Torr]$. This indicates that above a prefill gas pressure of 
$7 \times 10^{-5}[Torr]$ the given loop voltage is not sufficient to achieve the critical degree of ionization, $\gamma_{iz}(t_{RIB})$, as shown in Figure \ref{Figure4}(b). That is, the maximum prefill gas pressure available for plasma burn-through with the given $20 [V]$ loop voltage exists between $5 \times 10^{-5}[Torr]$ and $7 \times 10^{-5}[Torr]$.


For the low prefill gas pressure cases, the corresponding peak values of $P_{rad+iz}$ in Figure \ref{Figure4}(c) indicate the RIB. As shown in Figure \ref{Figure4}(c), ${P}_{iz+rad}$ is dominant in $P_{Loss}$ during the plasma burn-through phase, and the peak of $P_{Loss}$ coincides with the RIBs. Therefore, the required electric field for plasma burn-through is mainly determined by the ${P}_{iz+rad}$. It should be noted that the RIB rises as prefill gas pressure increases in Figure \ref{Figure4}(c). This is due to the fact that neutrals are strong energy sinks. That is, the larger number of neutrals at a high prefill gas pressure results in higher ${P}_{rad}$. In addition, there are more neutrals to be ionized at a high prefill gas pressure, thereby increasing ${P}_{iz}$. 

The increase in RIB can also be seen in \Eref{Pizrad}. \Eref{Pizrad} indicates that the maximum $P_{rad+iz}(t_{RIB})$ is
\begin{eqnarray}
P_{rad+iz}(t_{RIB}) = \frac{V_V \mathcal{P}_{RI}(T_e) n_D^0(0)^2}{4}. \label{RIB_magnitude}
\end{eqnarray}
$n_D^0(0)$ is proportional to the prefill gas pressure $p$. Hence, $P_{rad+iz}(t_{RIB})$ also increases proportionally with the square of $p$, if $T_e$ is identical at the $t_{RIB}$. Since $T_e$ during the plasma burn-through phase does not vary significantly, \Eref{RIB_magnitude} is another indication for the increase in RIB with prefill gas pressure.

The required electric field to overcome the RIB, $E_{RIB}$, can also be calculated using \Eref{RIB_magnitude}, i.e.  $P_{Oh} = P_{rad+iz}(t_{RIB}) $. Since the ohmic heating power $P_{Oh}$ is $V_p (E^2/\eta_s)$, the required electric field $E_{RIB}$ is 
\begin{eqnarray}
E_{RIB}^2 = \frac{\eta_s V_V \mathcal{P}_{RI}(T_e) n_D^0(0)^2}{4V_p}  \nonumber \\
E_{RIB}=0.011\sqrt{\frac{V_V  \mathcal{P}_{RI}(T_e)}{V_p T_e^{3/2}}} \times n_D^0(0)   \label{burn-through_criterion}
\end{eqnarray}
where $\eta_s$ is Spitzer resistivity, i.e. $\eta_s = 5 \times 10^{-4} \times T_e^{-3/2} [eV]$\cite{wesson}. 


\section{Review of \textcolor{correction}{wall-sputtering} models for carbon wall}\label{ReviewofPSImodelsforcarbonwall}

In the previous section, the burn-through process of a pure deuterium plasma was described. Now, impurity effects are taken into account. 

 In the carbon wall, chemical sputtering, emitting hydrocarbon molecules such as $CD_4$, is dominant \textcolor{correction}{if $T_e$ is lower than 100 eV, which is the case for the plasma burn-through phase.} Since the chemical sputtering yield is weakly dependent on the incident ion energy, the carbon sputtering yield due to deuterium ion bombardment is assumed to be a constant $0.03$, based on the experimental data from laboratory plasmas\cite{Mech1998}. In the case of oxygen ion bombardment, most oxygen ions are recycled as a carbon monoxide, $CO$, at a carbon target \cite{Davis1997}. According to this, oxygen sputtering is also modelled for the carbon wall JET with a constant sputtering yield, $1$. The details of the chemical sputtering model used are given in \cite{DYON}. \cor{The DYON simulation results using \textcolor{correction}{the wall-sputtering model} for the carbon wall have shown good agreement with the experimental data of JET with the carbon wall as presented in \cite{DYON}. However, it should be noted that the chemical sputtering model used was a simplified model, without considering some issues such as chemical sputtering by neutrals, physical sputtering, and a-CH film formation on the wall surface. These can increase the carbon sputtering yield. }  

\begin{figure}[H]
\begin{center}
\includegraphics[width=1\textwidth]{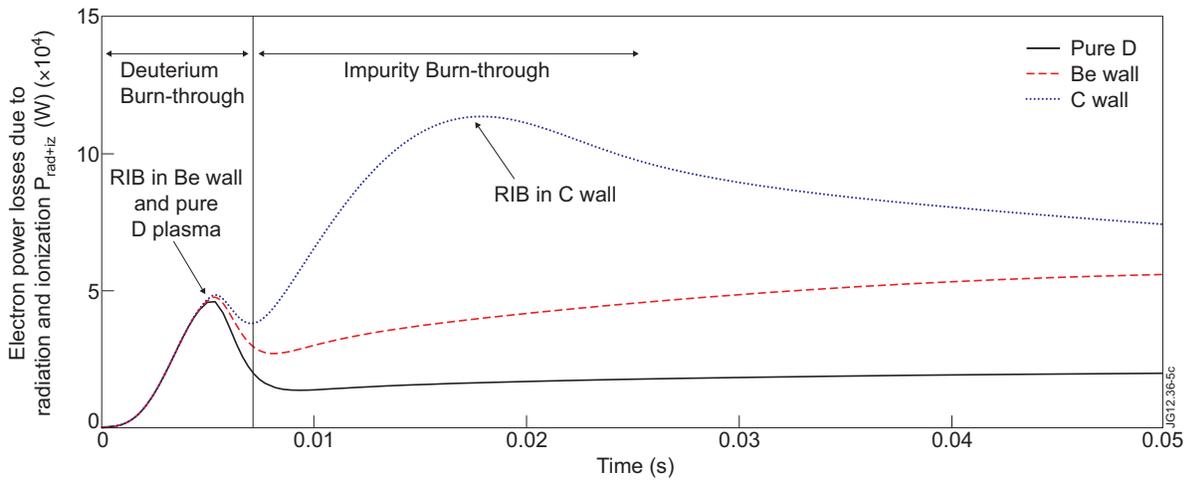} 
\caption{DYON simulation results for a pure deuterium plasma or with \textcolor{correction}{wall-sputtering} models i.e. carbon wall or beryllium wall. Each line indicates the electron power losses due to the radiation and ionization: carbon wall (dotted blue), beryllium wall (dashed red), and pure deuterium plasma (solid black). In the case of the carbon wall, the first peak (mainly from deuterium radiation) is much smaller than the second peak, which results from carbon impurities.} \label{Figure5}
\end{center}
\end{figure}

One of the important features with the carbon wall is that the electron power loss is dominated by the carbon burn-through. Figure \ref{Figure5} shows DYON simulation results of the electron power losses due to radiation and ionization with different wall models at JET : with a carbon wall (dotted blue), with a beryllium wall (dashed red), or for a pure deuterium plasma (solid black).  In the case of the carbon wall in Figure \ref{Figure5}, the second peak of $P_{rad+iz}$ represents the RIB required to be overcome for carbon burn-through. As shown, the RIB for carbon burn-through is critical for $I_p$ ramp-up as it is much higher than the RIB for deuterium burn-through. It should be noted that such a large RIB does not appear for beryllium burn-through, computed with the \textcolor{correction}{modified wall-sputtering} model which will be discussed in the next section. 

 

\section{\textcolor{correction}{Wall-sputtering} models for ITER-like wall}\label{NewPSImodelsforITER-Likewall}


One of the main differences of the beryllium wall compared to the carbon wall is that physical sputtering is dominant due to its low threshold energy\cite{Beryllium}. It is well known that a physical sputtering yield is a function of incident ion energy. The Bohdansky formulae for physical sputtering yield has been given as \cite{sputteringformula},\cite{surfacebindingenergy}
\begin{eqnarray}
Y=Q \times S_n(\epsilon) \times g(\delta) \label{Bohdansky}
\end{eqnarray}
where $Q$ is yield factor and $S_n$ is nuclear stopping cross-section which is given by
\begin{eqnarray}
S_n = \frac{3.441 \sqrt{\epsilon} \ln (\epsilon+2.718) }{1+6.355\sqrt{\epsilon}+\epsilon(6.882\sqrt{\epsilon}-1.708)}. \label{nuclearstopping}
\end{eqnarray}
$\epsilon$ is defined to be $E_0 / E_{TF}$ where $E_0$ and $E_{TF}$ are the ion wall-impacting energy and the Thomas-Fermi energy, respectively\cite{sputteringformula},\cite{surfacebindingenergy}. \textcolor{correction}{Assuming a typical sheath formation of negative potential at the wall \cite{stangeby},} the ion wall-impacting energy $E_0$ can be calculated as $2kT_i+3kT_e$. That is, the ions have $2kT_i$ of thermal energy when entering the sheath edge (i.e. ion heat tranmission coefficient $\gamma_i = 2$), and ion's energy gain within a sheath is approximately $3kT_e$. The function $g(\delta)$ is defined to be
\begin{eqnarray}
g(\delta)=(1-\delta^{2/3})(1-\delta)^2, \label{gfunction}
\end{eqnarray}
where $\delta$ is defined to be $E_{th} / E_0$\cite{stangeby}. $E_{th}$ indicates the threshold energy for physical sputtering. 

\begin{figure}[H]
\begin{center}
\includegraphics[width=1\textwidth]{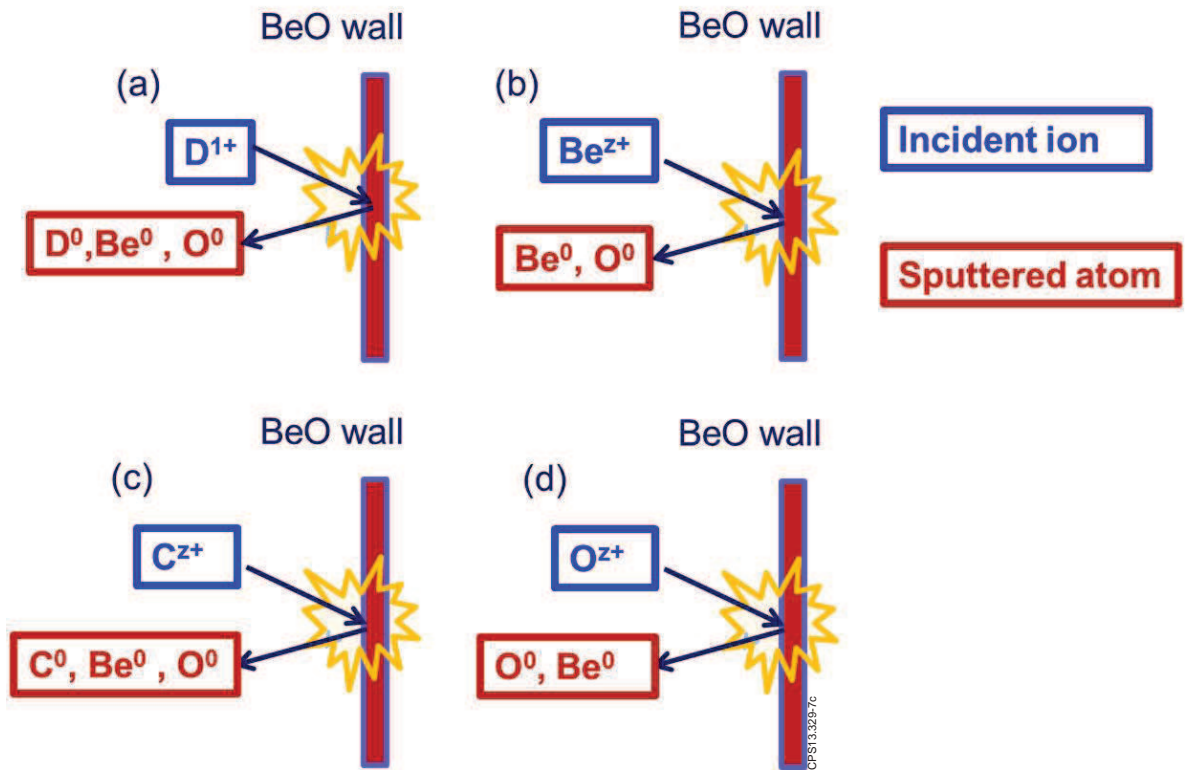}
\caption{\textcolor{correction}{Wall-sputtering and recycling} models used in DYON simulations with the ITER-like wall }  \label{picture_PSI_Be}
\end{center}
\end{figure}

\cor{Although there was a small air leak ($\sim 1.5 \times 10^{-6}$ $[Torr \ m^3 /  sec]$) in JET during the 2011/2012 experimental campaigns with the ITER-like wall, the oxygen level in the residual gas in the vacuum vessel remained lower than in JET with the carbon wall \cite{Kogut2012}. This implies the oxygen in the air forms a BeO monolayer on the wall.} Figure \ref{picture_PSI_Be} describes the \textcolor{correction}{wall-sputtering} models used in the DYON simulation for the ITER-like wall. Deuterium, carbon, oxygen,  and beryllium ions are modelled to be incident ions on the BeO wall (or pure Be wall), \textcolor{correction}{resulting in Be (and O) sputtering due to the wall erosion. Since the threshold energy of tungsten divertor significantly exceeds the range of  incident ion energy during the burn-through phase, (e.g. $E_{th}$ of tungsten is $220 [eV]$ with D ion bombardment. \cite{stangeby}), tungsten sputtering is not taken into account here. The sputtering yield is subject to the oblique angle of incident ions. However, it is difficult to find the effective oblique incidence angle for 0D simulations. In addition, a rapid evolution of the field line angles or even the magnetic geometry during plasma formation from open to closed field lines makes it difficult to assess the incidence angle. For a starting point of this study, we assume normal incidence. In this case, the Bohdansky formulae agrees well with more sophisticated models, such as TRIM \cite{Stamp_TRIM}.}

\textcolor{correction}{\cor{The BeO layer contacting the plasma (mainly at the limiter area) would be eroded by the plasma.} It is assumed that the surface of the Be wall is oxidized first, and the BeO layer is removed by ion bombardments after a certain erosion period, changing the BeO wall to the pure Be wall. Including the effects of the BeO layer erosion, the impurity influx due to the physical sputtering is modelled as 
\begin{eqnarray}
\Gamma^0_{Be,in}=V_p \sum_A \sum_{z \ge 1} (C_{BeO} Y^{BeO}_A + (1-C_{BeO})Y^{Be}_A) \frac{n^{z+}_A}{\tau_p} \\
\Gamma^0_{O,in}=V_p \sum_A \sum_{z \ge 1} C_{BeO} Y^{BeO}_A  \frac{n^{z+}_A}{\tau_p} \\
\Gamma^0_{D,in}=V_p Y^{D}_D  \frac{n^{1+}_D}{\tau_p} \\
\Gamma^0_{C,in}=V_p \sum_{z\ge 1}Y^{D}_D  \frac{n^{z+}_C}{\tau_p}
\end{eqnarray}
where superscript and subscript of $Y$ indicate the sputtered (or recycled) species and an incident ion, respectively. For example, $Y^{BeO}_D$ is BeO sputtering yield due to D ion bombardment. $C_{BeO}$ is defined as the BeO erosion coefficient, which is used to model the transition from BeO wall to Be wall at the end of the BeO layer erosion time $\tau_{erosion}$.  $C_{BeO}$ is 1 before $\tau_{erosion}$ [msec] and decreases to 0 after $\tau_{erosion}$. $\tau_{erosion}$ is adjusted to be 60 [msec] based on obtaining agreement of the synthetic data for bolometry and $Be^{1+}$ PM tube signals against the measured data.} The parameters required to calculate physical sputtering yields of the BeO wall (or pure Be wall) are given in Table \ref{BeSputteringParameters}. 

The incident ions are also recycled as neutrals at the wall with a fraction, called the recycling coefficient. It is observed at JET that during the plasma burn-through phase the carbon wall releases deuterium into the plasma, while the beryllium wall pumps them from the plasma\cite{Peter_2012IAEA}.  Hence, for DYON simulations with the carbon wall, an exponential decay model was used \textcolor{correction}{i.e. $Y^D_D = 1.1 \to 1$.}  However, the deuterium recycling coefficient with the ITER-like wall is modelled to grow and approach $1$ during the plasma burn-through phase. That is,  the exponential growing model is used for DYON simulations with the ITER-like wall \textcolor{correction}{i.e. $Y^D_D = 0.9 \to 1$. This is consistent with the outgassing observed after discharges with the ITER-like wall \cite{Philipps2013}.}  


\section{ITER-like wall simulations and comparison to the JET data}\label{ILWsimulationsandcomparisontotheJETdata}

\subsection{Validation of the new models}\label{Validationofthenewmodels}
\begin{figure}[H]
\includegraphics[width=1\textwidth]{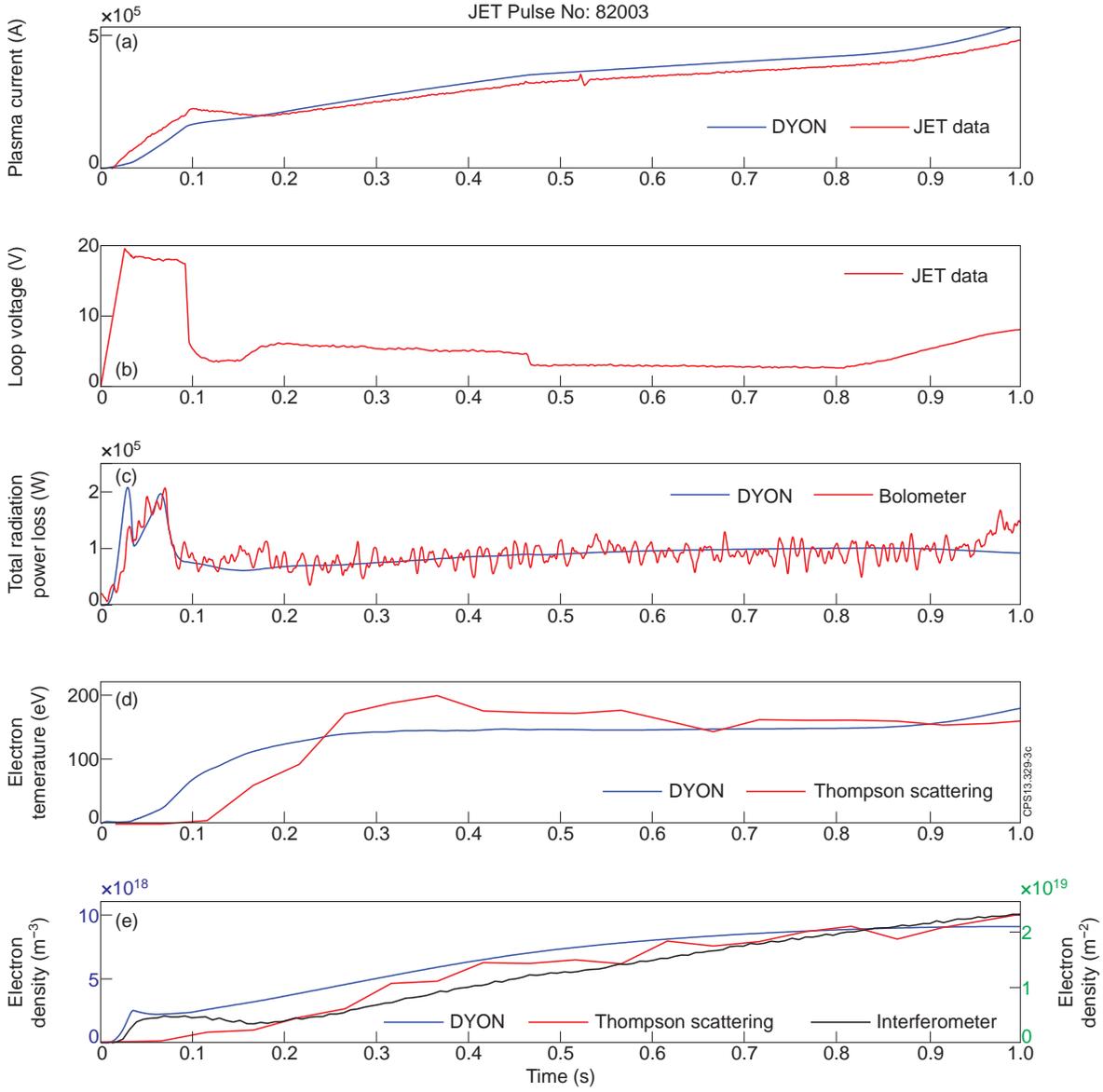}
\caption{DYON simulation results with the new models for the ITER-like wall and JET data. (a) Plasma current, (b) Loop voltage, (c) Total radiation power (Bolometry), (d) Electron temperature (Thomson scattering), (e) Electron density (Thomson scattering and Interferometry). The red lines (and the black line in (e)) indicate JET data for $\# 82003$, and the blue lines are the corresponding DYON simulation results. The  the condition given for the simulations is in Table \ref{conditionforILWsim}.}    \label{Demo82003a}
\end{figure}

The recent installation of the ITER-like wall at JET enables \cor{us} to validate the new sputtering (or recycling) models at the beryllium wall. For the validation of the new models, one typical discharge in JET has been selected  ($\# 82003$), and compared to the DYON simulation results. Figure \ref{Demo82003a} shows the DYON simulation results and the JET data with the ITER-like wall. The required input parameters for the DYON simulation, i.e. prefill gas pressure, loop voltage, plasma major and minor radius, toroidal magnetic field, and additional fuelling, are obtained from the measured data. The parameters given to perform the simulation are summarized in Table \ref{conditionforILWsim}.

The plasma current in the DYON simulation is in good agreement with the JET data. As shown in Figure \ref{Demo82003a}(b), the toroidal loop voltage decreases abruplty at 0.1 second due to the pre-programed use of a switching network to reduce the voltage from the ohmic transformer, a typical operation scenario at JET. This results in the sharp decrease in the $I_p$ ramp-up rate around 0.1 second in Figure \ref{Demo82003a}(a). The measured loop voltage is used as an input for the simulations. 

Figure \ref{Demo82003a}(c) shows the total radiation power loss.  \textcolor{correction}{In the simulations, the temporal behaviour of the radiation barrier is used to adjust $\tau_{erosion}$. The synthetic bolometry data can be well reproduced with  $\tau_{erosion}$(=60 [msec]). For this simulation, the initial carbon content  $n_C^0(0)$ is assumed to be $0.5 \%$ of prefill deuterium atoms $n_D^0(0)$. As will be discussed later, the assumption of  $n_C^0(0)$ has a small contribution to  the magnitude of the radiation barrier. Also, both $\tau_{erosion}$ and $n_C^0(0)$ do not have a significant influence on the evolution of other plasma parameters i.e. $I_p(t), T_e(t)$, and $n_e(t)$.}

The $T_e$ and $n_e$ in the simulations and the Thomson Scattering data approach similar values, but they have a discrepancy before 0.2 [sec]. The discrepancy is due to the limitations of the Thomson Scattering diagnostic, which can have significant error bars during the low density phase such as the plasma burn-through phase. The interferometry data show a better agreement with the density in the simulations during this early phase, as shown in Figure \ref{Demo82003a}(e). 


As ionization of deuterium and impurities proceeds, the photomultiplier tube data has a peak for each specific line emission, which is emitted from the deuterium atom or impurity ion in a \textcolor{correction}{certain} charge state.  Synthetic data of photomultiplier tube \cor{$I_{A^{z+}}$} can be calculated with the plasma parameters \cor{($n_e, n_A^{z+}$, and $T_e$)} obtained by the DYON simulation as
\begin{eqnarray}
I_{A^{z+}} = n_e n_A^{z+} \mathcal{PEC}_{A^{z+}}(n_e,T_e) [p ~ m^{-3} sec^{-1}] \label{syntheticPMtube}
\end{eqnarray}
where $\mathcal{PEC}_{A^{z+}}(n_e,T_e)$ is a photon emissivity coefficient, which is a function of electron density and temperature. Here, $A$ and $z+$ indicate the corresponding particle species and the charge state, respectively. The photon emissivity coefficients are obtained from the ADAS package \cite{Summers}. The \textcolor{correction}{synthetic volume emission} of $Be^{1+}(527 [nm])$,  $D^{0}(D alpha)$, and $C^{2+}(465 [nm])$ are compared with the measured \textcolor{correction}{photon flux} in Figure \ref{Demo82003b}(a)(b)(c). The temporal behaviour of the peaks in the synthetic data is coincident with the measured data. This implies that the ionization process of deuterium and impurities is reproduced correctly in the DYON simulations.   

\subsection{Beryllium sputtering in ITER-like wall}\label{sectionBesputteringinILW}

\textcolor{correction}{In the DYON simulations, the first radation peak for $Be^{1+}$ is reproduced showing a very good temporal agreement with the photomultiplier tube data in Figure \ref{BeSputteringModel}(a). However, the first radiation peak is removed when the initial beryllium content is not assumed in the simulations, as shown in Figure \ref{BeSputteringModel}(b). This implies that the first radiation peak results from the initial beryllium content rather than physical sputtering.  These particles are probably the beryllium atoms bonded weakly at the wall due to the migrations of beryllium in the vacuum vessel during previous experiments.}

\begin{figure}[H]
\begin{center}
\includegraphics[width=1\textwidth]{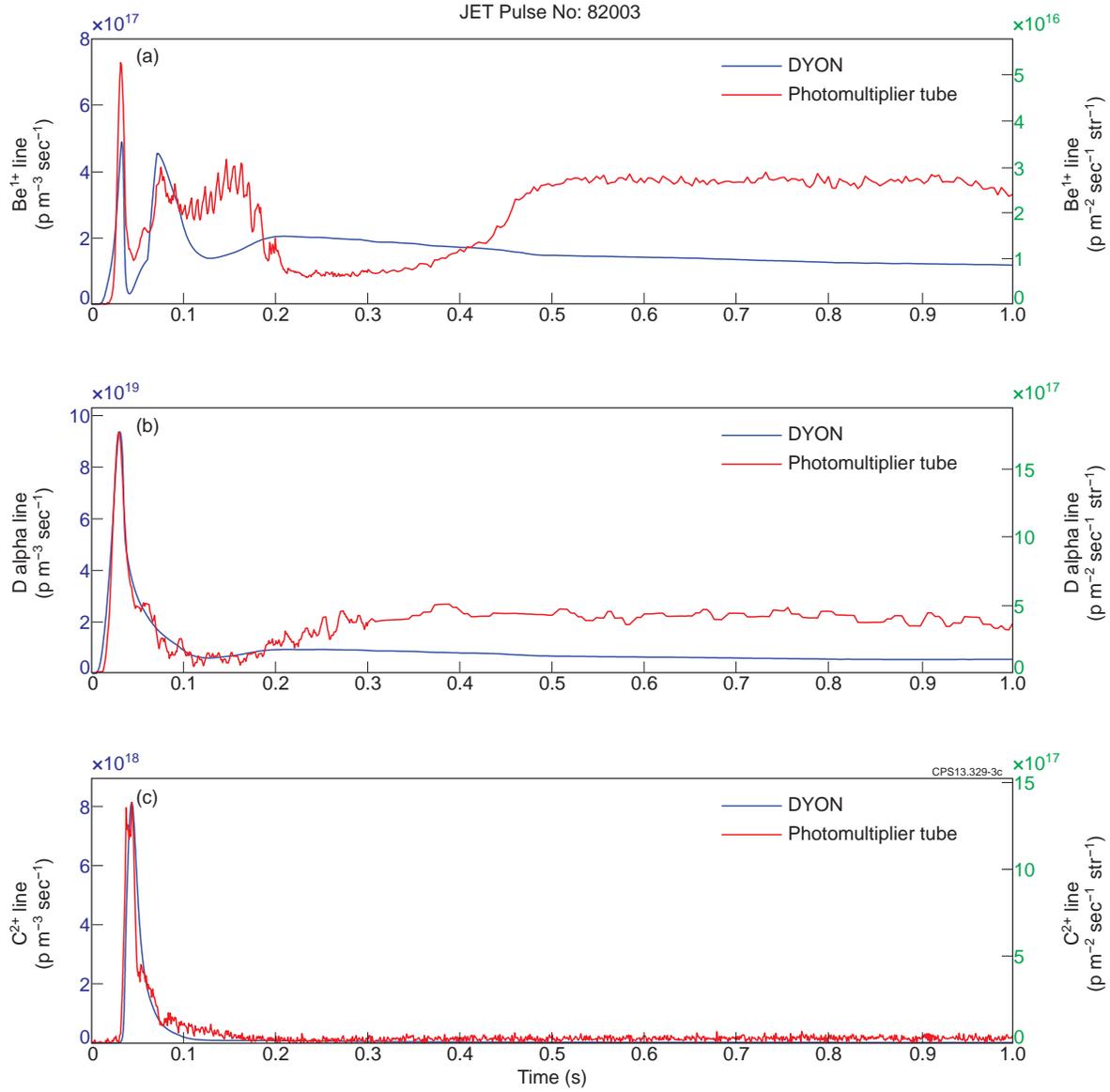}
\caption{The measured photomultiplier tube data and the synthetic photomultiplier tube data: (a) number of photons emitted from $Be^{1+}$ ($527 [nm]$), (b) number of photons emitted from $D^{0}$(D alpha), and (c) number of photons emitted from $C^{2+}$ ($465[nm]$). The red lines are the photomultiplier tube data in JET for $\# 82003$, and the blue lines are the synthetic data, calculated by DYON simulations. }  \label{Demo82003b}
\end{center}
\end{figure}

\begin{figure}[H]
\begin{center}
\includegraphics[width=1\textwidth]{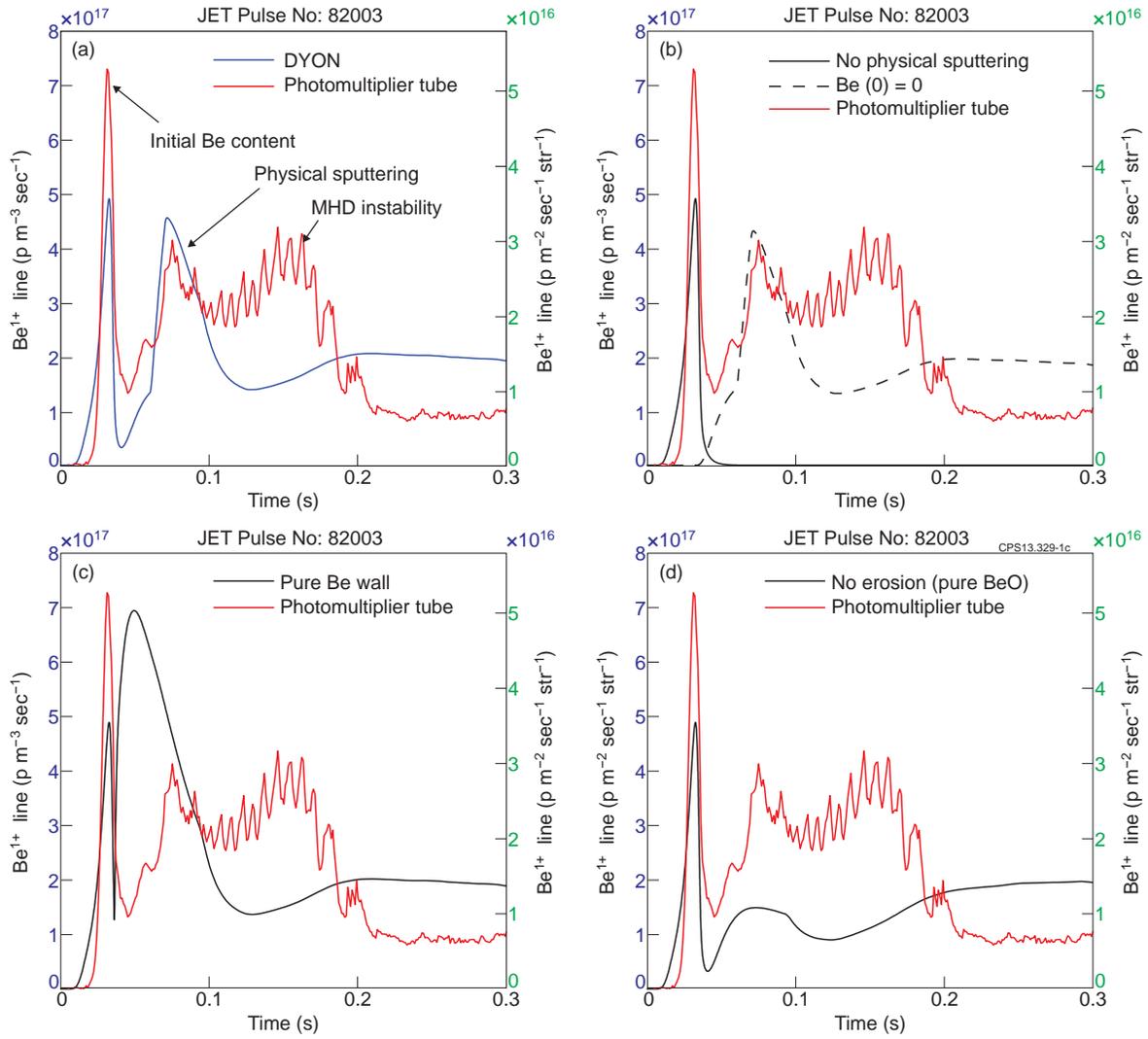}
\caption{\textcolor{correction}{These figures compare the photomultiplier tube data ($Be^{1+}$ ($527 [nm]$)) between the measured data and the synthetic data, and show the significance of the new models used in the DYON simulations. The red solid lines are the photomultiplier tube data in JET  for $\# 82003$. The blue solid line in (a) is the synthetic data with the condition given in Table \ref{conditionforILWsim} (i.e. with physical sputtering model on BeO layer, erosion model of BeO layer, and initial Be content). The black dashed line and solid line in (b) are without initial Be content or any physical sputtering model, respectively. The black solid line in (c) is for pure Be wall. The black solid line in (d) is without erosion model of BeO layer i.e. continuous BeO layer.} }  \label{BeSputteringModel}
\end{center}
\end{figure}

The secondary radiation peak in the photomultiplier tube data for $Be^{1+}$ is observed at around 0.07 [sec], as shown in Figure \ref{Demo82003b}(a).  It should be noted that such a secondary radiation peak does not appear for $D^{0}$ and $C^{2+}$ in Figure \ref{Demo82003b}(b)(c).  This is because the secondary radiation peak results from the wall-sputtering rather than the initial impurity content.  Wall-sputtering can occur only if the incident ion energy exceeds the threshold energy. This implies that the beryllium sputtering is delayed until the wall-impacting energy, $2kT_i+3kT_e$, exceeds the threshold energy. The secondary radiation peak is reproduced with the physical sputtering model, showing a good temporal agreement in Figure \ref{BeSputteringModel}(a).  Figure \ref{BeSputteringModel}(b) shows the DYON simulation results without the physical sputtering model. The secondary radiation peak does not appear without the physical sputtering model. This leads to the conclusion that  the secondary radiation peak results from the delayed physical sputtering. 

It is observed in many laboratory plasmas that the surface of beryllium tiles are easily oxidized \cite{Roth1997},\cite{Hirai},\cite{Lungu}. The high affinity of Be to O tends to result in the BeO layer on the wall, which has higher surface binding energy than the pure beryllium wall\cite{Roth1997}. In Table \ref{BeSputteringParameters}, the threshold energy of a deuterium incident ion for the physical sputtering on the BeO layer is $29[eV]$, which is much higher than that on a pure beryllium wall, $10[eV]$. The higher surface binding energy in the BeO layer makes the physical sputtering more difficult than in the pure beryllium wall. The simulation results using the different parameters for BeO and Be are compared in Figure \ref{BeSputteringModel}(a)(c). The secondary radiation peak in Figure \ref{BeSputteringModel}(c) occurs much earlier, showing a deviation from the measured value. 

\textcolor{correction}{Due to the erosion of BeO layer, the BeO sputtering model is switched to the Be sputtering at $\tau_{erosion} (\approx 60[msec])$. \Fref{BeSputteringModel}(d) shows BeO sputtering without switching to Be sputtering i.e. no BeO erosion model. The secondary peak shows a reasonable temporal agreement with the measured, but it decreases slowly, without the sharp peak observed in the measured and the simulated with the erosion model. Moreover, in this case the radiation resulting from oxygen would be much higher than measured.}

Based on the frequency of the fluctuating radiation, it is found that the third peak around $100 \sim 200[msec]$ in the photomultiplier tube data for $Be^{1+}$ is probably due to an MHD instability in the plasma, which is not modelled in the DYON simulation. Since a detailed investigation on the MHD instabilities is beyond the scope of this paper, this is not discussed further here.


\subsection{Deuterium recycling coefficient} \label{SectionDeuteriumrecyclingcoefficient}

The initial deuterium recycling coefficient $Y^D_D(0)$ is subject to the wall condition, and $Y^D_D(t)$ approaches 1 \textcolor{correction}{as the deuterium inventory at the wall saturates.} The varying deuterium recycling coefficient $Y^D_D(t)$ can be modelled by\cite{DYON}
\begin{eqnarray}
Y^D_D(t) =c_1 - c_2(1-exp(-\frac{t}{c_3})) \label{Drecmodel}
\end{eqnarray}
$Y^D_D(t)$ is adjusted by the combination of constants $c_1$, $c_2$, and $c_3$. In order to test whether $Y^D_D$ is growing or decaying during the plasma burn-through phase, the constants are assumed to be $c_1 = 0.9$, $c_2 = -0.1$, and $c_3 = 0.1$ for the growing model, and $c_1 = 1.1$, $c_2 = 0.1$, and $c_3 = 0.1$ for the decay model. 

\begin{figure}[H]
\begin{center}
\includegraphics[width=1\textwidth]{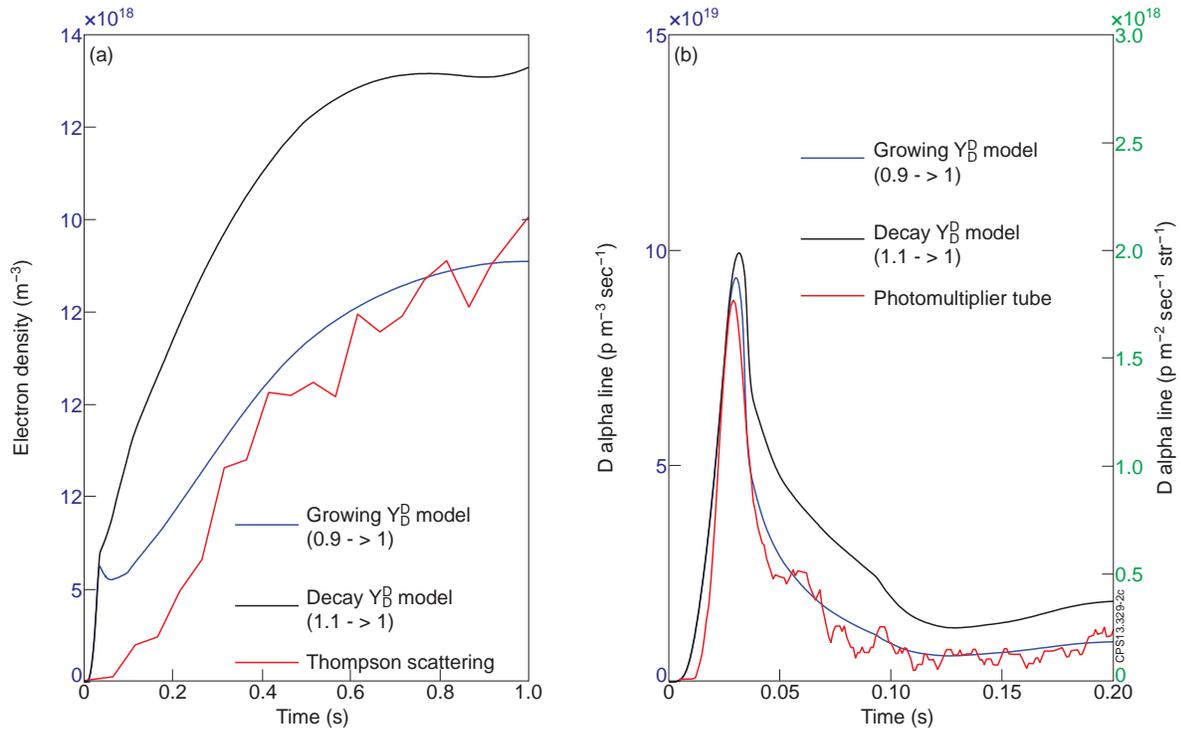}
\caption{\textcolor{correction}{The figures show the effects of the deuterium recycling coefficients. (a) electron density (b) number of photons emitted from $D^{0}$(D alpha) ($465[nm]$). The red lines are (a) Thomson scattering data (b) the measured photomultiplier tube data in JET ($\# 82003$).  The blue and black lines are the DYON simulation results with the growing model or decay model of $Y^D_D$, respectively.} }  \label{Scanning_Drecycling_82003a}
\end{center}
\end{figure}

Figure \ref{Scanning_Drecycling_82003a} shows the different simulation results between the two sets of parameters used. The decay model implies additional release of deuterium atoms during the plasma burn-through phase. This results in a too high electron density in the simulation compared to the Thomson scattering data in Figure \ref{Scanning_Drecycling_82003a}(a).  The discrepancy is reduced by the growing model as indicated with the blue solid lines in Figure \ref{Scanning_Drecycling_82003a}(a). This implies that some portion of incident deuterium ions are retained at the wall rather than being recycled. Based on this, the growing model for recycling is used for the simulations with the ITER-like wall in Figure \ref{Demo82003a}. It should be noted that for the carbon wall simulations the DYON simulation results with the decay model showed better agreement with the JET data.\cite{DYON}. 


Figure \ref{Scanning_Drecycling_82003a}(b) gives another indication that the growing model is required for the ITER-like wall. The synthetic data for the D alpha line using the growing model shows very good agreement against the measured value for $0 \sim 0.2 [sec]$. However, with the decay model, the synthetic data deviates significantly from the measured value as shown in Figure \ref{Scanning_Drecycling_82003a}(b). This gives a confidence on the fact that such additional release of deuterium atoms during the plasma burn-through phase is not probable in the ITER-like wall.

\subsection{Gas fuelling}\label{Gasfuelling}

\begin{figure}[H]
\begin{center}
\includegraphics[width=1\textwidth]{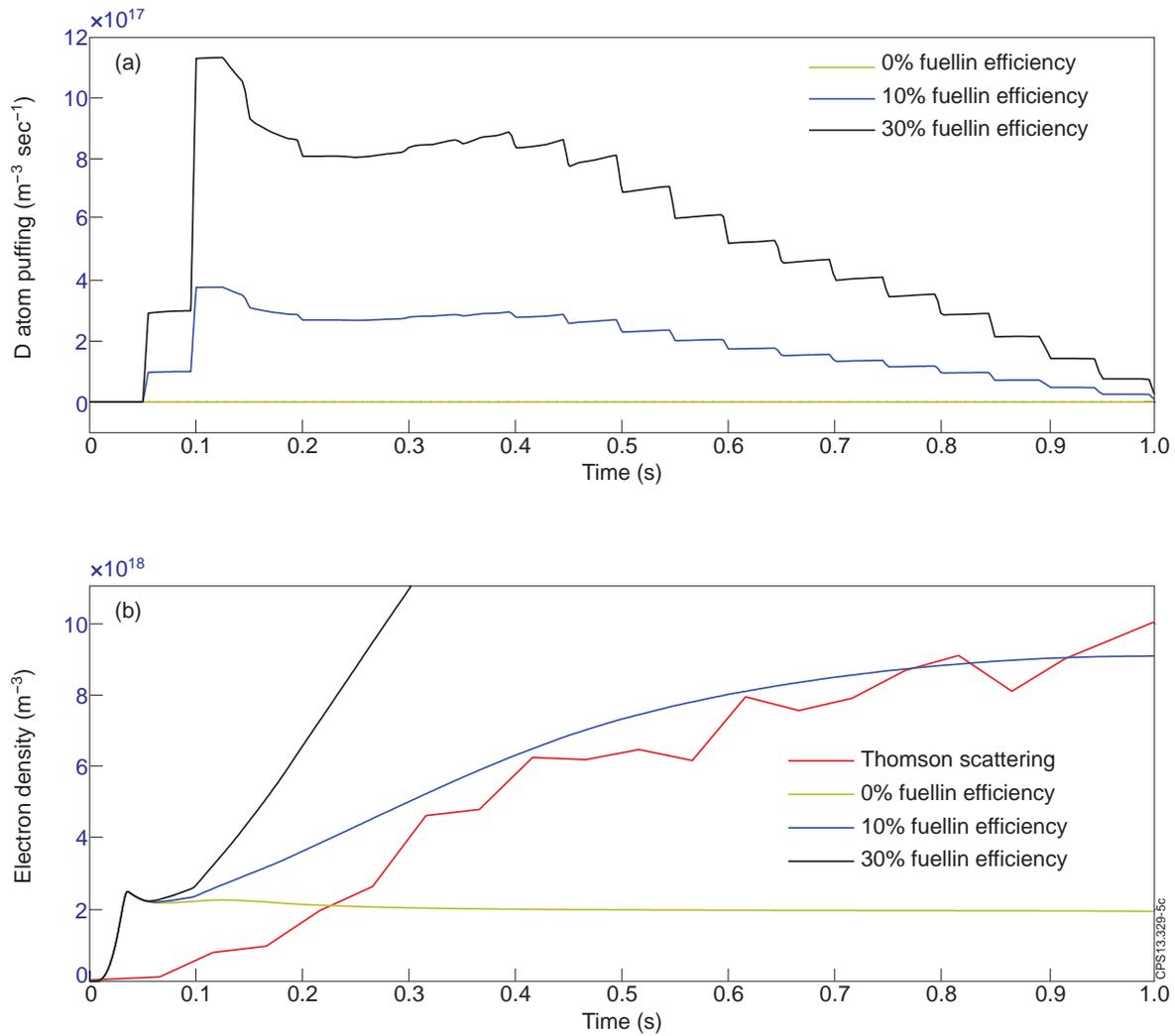}
\caption{\textcolor{correction}{The figures show the effects of fuelling efficiency. (a) D atom puffing with the assumed fuelling efficiency (black $30\%$, blue $10\%$, and green $0\%$), (b) electron densities, obtained by DYON simulation (black $30\%$, blue $10\%$, and green $0\%$) and measured by Thomson scattering (red) in JET ($\# 82003$).}}  \label{Scanning_fuelling_82003a}
\end{center}
\end{figure}

One of the main differences in the JET operation scenario with the ITER-like wall is the use of additional gas fuelling around $0.1 [sec]$. This is to compensate for the gas pumping effect at the wall as seen with the new wall\cite{Peter_2012IAEA}. The additional gas fuelling influences on the particle balance of deuterium atoms\cite{DYON},  
\begin{eqnarray}
\fl \frac{dn^0_D}{dt}= \frac{1}{\gamma_n^D V_V} ( V_p \mathcal{R}^{1+\rightarrow 0}_{D,rec}n_e n^{1+}_D \nonumber \\ 
- V^D_n \mathcal{R}^{0 \rightarrow 1+}_{D,iz}n_e n^0_D  - V^D_n\sum_I \sum_{z \geq 1} \mathcal{R}^{z+\rightarrow (z-1)+}_{I,cx}n^0_Dn^{z+}_I) + \frac{\Gamma_{D,in}^{total}}{\gamma_n^D V_V},  \label{DeuteriumAtomPtlBalance}
\end{eqnarray}
where the total influx of deuterium atoms $\Gamma_{D,in}^{total}$ is
\begin{eqnarray}
\Gamma_{D,in}^{total} = V_p \frac{Y^D_D n^{1+}_D}{\tau_D} + \Gamma^{eff}_{D,in}. \label{TotalInfluxofDatom}
\end{eqnarray}
The first term in \Eref{TotalInfluxofDatom} is the influx of the deuterium atoms recycled at the wall, and the second term is the additional gas fuelling $\Gamma^{eff}_{D,in}$, which was not included for the carbon wall simulations. In the simulations for the ITER-like wall, 
$\Gamma^{eff}_{D,in}$ is modelled as 
\begin{eqnarray}
\Gamma^{eff}_{D,in} = \psi_{puffing} \Gamma^{GIM}_{D,in},  \label{AdditionalGasPuffing}
\end{eqnarray}
where $\Gamma^{GIM}_{D,in}$ is obtained from the data of the Gas Injection Modules in JET\cite{GIMinJET} and $\psi_{puffing}$ is a fuelling efficiency. \textcolor{correction}{The gas injection module used for the fuelling at $0.1 [sec]$ is located on the top of Oct 8 in JET, and the 4 main pumping ports, 2 Neutral Beam Injection (NBI) ports, and 1 Lower Hybrid (LH) ports are on the outer midplane of the vessel.} Since a significant fraction of the injected gas is pumped out immediately rather than being ionized, the fuelling efficiency must be evaluated for the effective influx of deuterium atoms. Figure \ref{Scanning_fuelling_82003a} shows the differences in the simulation results when assuming $30\%$ and $0\%$ for the fuelling efficiency $\psi_{puffing}$. Since gas injection is applied from $0.1[sec]$ as can be seen in \ref{Scanning_fuelling_82003a}(a), the simulation results does not show discrepancy until then. However, the electron density with $30\%$ fuelling efficiency increases excessively from $0.1[sec]$ onwards as shown in \Fref{Scanning_fuelling_82003a}(b).  In contrast, without the fuelling model i.e. $0 \%$ fuelling efficiency, the electron density in Figure \ref{Scanning_fuelling_82003a}(b) is much lower than Thomson scattering data. Scanning of the fuelling efficiency, it has been found that the simulation results with $10\%$ fuelling efficiency agree well with the JET data. Based on this, $10\%$ fuelling efficiency is assumed for the simulations in \Fref{Demo82003a} and \Fref{Demo82003b}.


\subsection{Initial carbon content}

Although all CFC tiles have been removed from the plasma facing components in the ITER-like wall, the $C^{2+}$ line emission is still observed in the ITER-like wall JET data as shown in Figure \ref{Demo82003b}(c). This requires the assumption of an initial carbon content $n^0_C(0)$ for the ITER-like wall simulations. 

\begin{figure}[H]
\begin{center}
\includegraphics[width=1\textwidth]{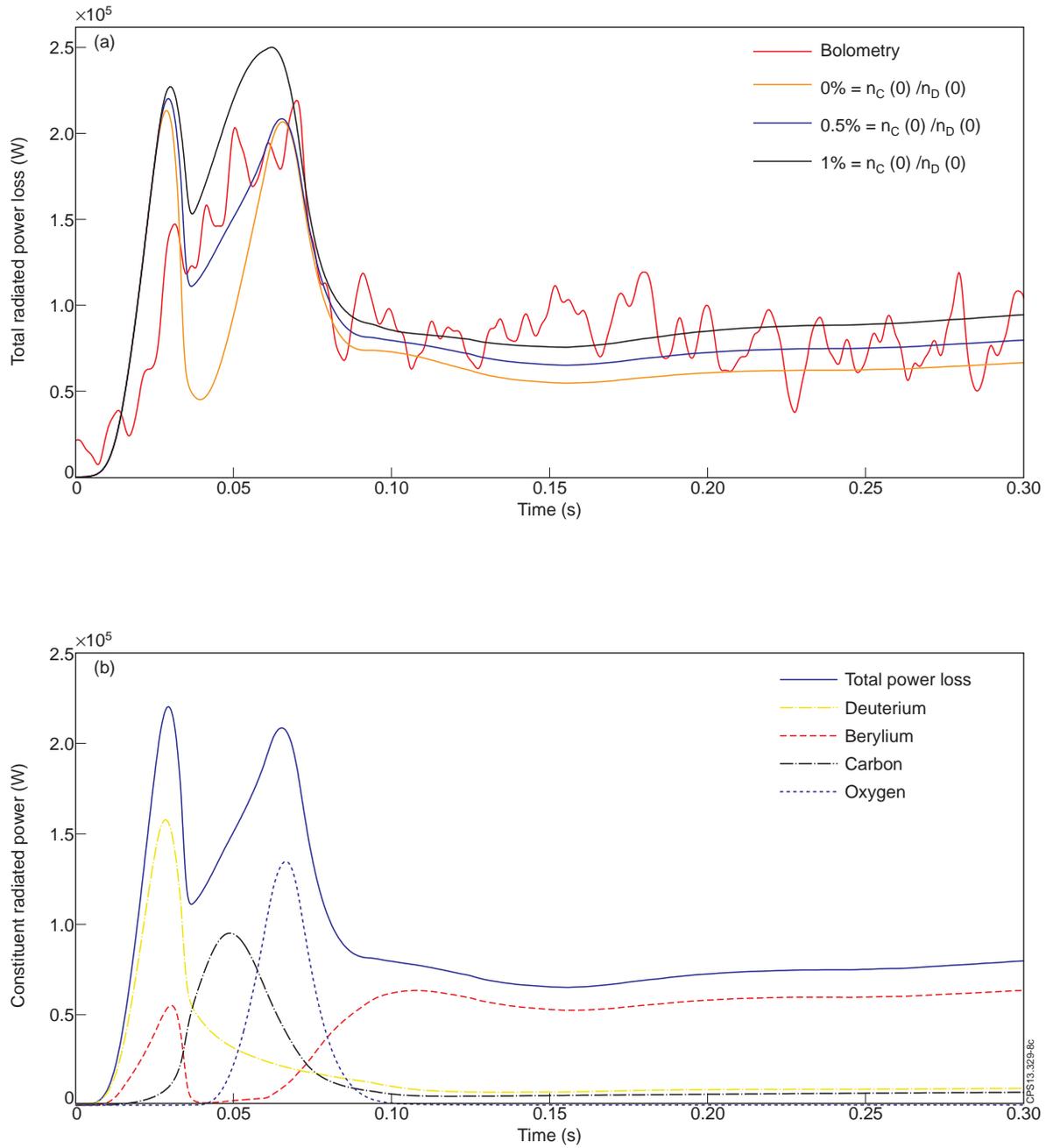}
\caption{\textcolor{correction}{The figures show the effects of $n^0_c(0)$ on the radiation barrier and the consitituent radiated power. (a) bolometry data (red) in JET ($\# 82003$), and the simulated radiated power in the DYON simulations (solid black $1 \%$, solid blue $0.5 \%$, and solid green $0 \%$ of $n^0_c(0)$). (b) the constituent radiated power (solid blue: total radiated power, dashed red: Be, dashed green: D, dashed black: C, dashed blue: O). For the simulation in (b), the $n^0_c(0)$ is assumed to be $0.5\%$, as given for the simulation in Figure \ref{Demo82003a} and \ref{Demo82003b}. } }  \label{Scanning_initialC_82003a}
\end{center}
\end{figure}

\cor{The total carbon content assumed at plasma initiation is between $0 \sim 1 \%$ of the prefill deuterium atom density $n_D^0(0)$; $n_C^0(0) \sim 10^{16}$ $[m^{-3}]$. This value is lower than the carbon content during the main heating phase of JET plasmas, which is reported \cite{Coenen2012} as $0.1 \sim 0.2 \%$ of the plasma density (in the range of $10^{19} \sim 10^{20}$ $[m^{-3}]$) during the heating phase.}
Figure \ref{Scanning_initialC_82003a}(a) shows the total radiation power loss in the simulations with  $n^0_C(0)$. Here, the initial carbon content $n^0_C(0)$ is assumed to be \textcolor{correction}{$0 \%$, $0.5 \%$ or $1 \%$} of the initial deuterium atom density $n^0_D(0)$, respectively. As shown in Figure \ref{Scanning_initialC_82003a}(a), without $n^0_C(0)$, the synthetic radiation barrier deviates from (below) the bolometry data.  The magnitude of the total radiation power loss in the simulation with \textcolor{correction}{$
1 \%$} $n^0_C(0)$  exceeds the peak of the bolometry data. However, in the case of the simulation with \textcolor{correction}{$0.5 \%$} $n^0_C(0)$, the radiation barrier shows good agreement. Based on this, \textcolor{correction}{$0.5 \%$} $n^0_C(0)$ is assumed in the simulations. 


Figure \ref{Scanning_initialC_82003a}(b) shows the consitituent radiated power losses in the case of the simulation with \textcolor{correction}{$0.5 \%$}  $n^0_C(0)$. It can be seen that the radiation is dominated in turn by the radiated power loss from deuterium, carbon, oxygen, and beryllium. The radiated power loss due to beryllium is not significant in the radiation barrier during the plasma burn-through phase. 

\section{Operation space for plasma burn-through in JET}\label{Operationspaceforplasmaburn-through}
\begin{figure}[H]
\begin{center}
\includegraphics[width=1\textwidth]{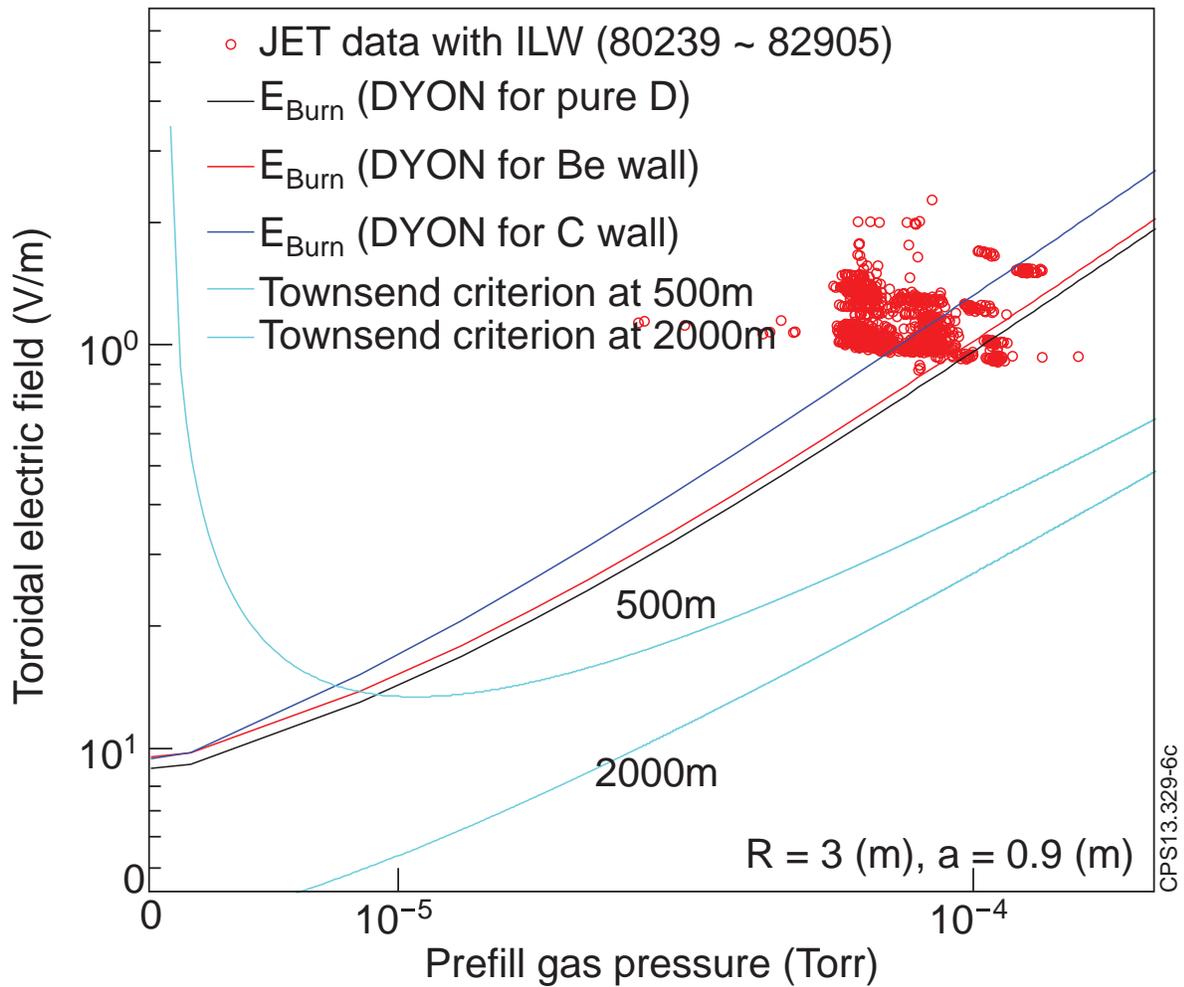}
\caption{The cyan lines show the Townsend criterion at different effective connection lengths, as indicated with $500$ and $2000 [m]$, respectively.  The black, red, and blue lines indicate the criterion for plasma burn-through, i.e. the minimum electric field for plasma burn-through in the case of a pure deuterium plasma (black), beryllium wall (red), and carbon wall (blue), respectively. The \textcolor{correction}{wall-sputtering} models described in section \ref{ReviewofPSImodelsforcarbonwall} and \ref{NewPSImodelsforITER-Likewall} are used for the simulations, and the required plasma parameters are given by Table \ref{defaultvalues}.  The area above both the burn-through criterion and Townsend criterion represents the operation space available for successful start-up in JET. \cor{The red circles indicate the successful plasma burn-through in JET experiments with ITER-like wall ($\# 80239 \sim 82905$).} }  \label{Figure6}
\end{center}
\end{figure}

Figure \ref{Figure6} compares the Townsend criterion and the criterion for plasma burn-through. The cyan lines in Figure \ref{Figure6} are the minimum electric field for electron avalanche $E_{avalanche}$ for \cor{two} effective connection lengths $L_f$ \cor{($=500$ and $2000$ $[m]$)} in JET, drawn analytically by using the Townsend criterion in \Eref{TownsendCriterion}. The Townsend criterion shows that there is an optimum range of the prefill gas pressure at which the lowest toroidal electric field is available for electron avalanche. However, for successful tokamak start-up, operation space is not only determined by the Townsend criterion, but also by the criterion for plasma burn-through. The black, red, and blue lines in Figure \ref{Figure6} represent the required electric field for plasma burn-through $E_{Burn}$ without or with impurities, obtained using DYON simulation results. The \textcolor{correction}{wall-sputtering} models described in section \ref{ReviewofPSImodelsforcarbonwall} and \ref{NewPSImodelsforITER-Likewall} are used for the DYON simulations. Plasma parameters assumed for the DYON simulation are indicated in Table \ref{defaultvalues}.  

The criterion for plasma burn-through is computed from the numerical simulations using several assumptions, subject to wall conditions and operation scenario. However, the simulation results provide informative insight on the operation space.  As shown in Figure \ref{Figure6}, the required electric field for plasma burn-through increases monotonically as prefill gas pressure rises since the RIB is greater at a high prefill gas pressure. This monotonic increase with prefill gas pressure is consistent with \Eref{burn-through_criterion}, which is analytically derived for deuterium burn-through.

If \textcolor{correction}{the effects of the impurities from the wall} are included, the required electric field for plasma burn-through increases, thereby reducing the operation space available. As can be seen in Figure \ref{Figure5}, the RIB for carbon burn-through is much greater than for deuterium burn-through. Hence, the required loop voltage in the carbon wall is significantly higher than for a pure deuterium plasma. This results in the smaller operation space available in the carbon wall as shown in Figure \ref{Figure6}. However, the RIB for beryllium burn-through is not significant in Figures \ref{Figure5}. With the beryllium wall, the critical RIB to be overcome is for deuterium rather than for beryllium \textcolor{correction}{as long as other impurities are not significant \cite{PSIusingDYON}.} This implies that lower loop voltage can be used for plasma burn-through in the ITER-like wall i.e. larger operation space available compared to the carbon wall as shown in Figure \ref{Figure6}. \cor{The red circles in Figure \ref{Figure6} show successful plasma burn-through in JET experiments with the ITER-like wall ($\# 80239 \sim \# 82905$). It should be noted that most successful shots are located above the red line, which is the criterion for plasma burn-through with the beryllium wall.}

The criterion for plasma burn-through in Figure \ref{Figure6} can change with different assumptions in the DYON code. That is, the increase in required loop voltage with increasing prefill gas pressure is steeper for higher recycling coefficient (or sputtering yield) or for higher ratio of the effective vessel volume to the plasma volume i.e. $V_V / V_p$. In tokamaks, the recycling coefficients and the sputtering yields vary due to the effects of deuterium retention and impurity migration in the wall.  In addition, $V_V / V_p$, which is related to dynamic neutral gas fuelling from the ex-plasma volume, is also varying according to the operation scenario. Hence, to find the precise operation space using the simulations, the information on the wall conditions and operation scenario in each shot should be specified. 

\section{Discussion}\label{sectionDiscussion}
The investigation of plasma burn-through has been published only with 0D simulations \cite{lloyd_ITER}\cite{DYON}\cite{TFTR_burn-through}\cite{scenplint}. Since the closed flux surfaces (CFSs) are not established yet during the plasma burn-through phase at low plasma current, a 2D approach of numerical simulation is extremely difficult. Fortunately, the results of DYON simulations, (also 0D), show good agreement with JET data. This implies the assumption of a uniform temperature and density in a numerical simulation is reasonable to compute the gross energy and particle balances during the plasma burn-through phase. Probably, this is due to the open field configurations during the plasma burn-through phase. With the open magnetic field lines, the parallel thermal conduction and particle diffusion would be significant. Quantitative investigations on the profile effects of temperature and density will be interesting to confirm this. \cor{Regarding the profile of plasma current, a flat current profile is assumed i.e. $l_i=0.5$. According to a scan of $l_i$ with the DYON code, the internal inductance does not have significant effect on the simulation results since most of the power is consumed by ohmic heating.}

\textcolor{correction}{The synthetic PM tube data is not dependent on the initial electron temperature $T_e(0)$ assumed in the simulations. $T_e(t)$ quickly saturates to the same value in a few millisecond regardless of the assumed value of $T_e(0)$. It has been checked in the simulations, starting with $T_e(0)$ = 1, 3, and 5 [eV]. }

\cor{We assume fully dissociated D gas, and this is a reasonable assumption for $T_e(0)=1$ $[eV]$, the initial condition in the simulation. The dissociation energy per D atom $P_{dis}$ is just  $2.26$ $[eV]$. Compared to D ionization energy $13.6$ $[eV]$, this is small.  Moreover, radiation power loss is much greater than D ionization energy i.e. $P_{rad}$ $>$ $P_{iz}$ $>$ $P_{dis}$.  This has been checked in the simulations by adding $2.26$ $[eV]$ to D ionization energy; the simulation results did not show any visible difference. Hence, the D dissociation energy can be ignored in the simulation.}

\textcolor{correction}{One of the underlying assumptions in the simulations is that all deuterium atoms are accessible to the plasma. \cor{This implies the vessel volume determines the neutral influx into the plasma. However, deuterium atoms in large ports for diagnotics or additional heating and near pumping ports are impeded to approach the plasma. For example, the neutral particles in the cryogenic pumping chamber for NBI system (i.e. $V_{NBI}$) cannot access to the main plasma. Hence, $V_{NBI}$ should not be included in $V_V$.} The initial peak of  bolometry data is proportional to the number of deuterium atoms in the vessel. Comparing the reproduced bolometry data with the measured, we could find the effective vessel volume $V_V$ is around 100 $[m^3]$ in JET, although the total volume of the vacuum vessel is 189 $[m^3]$.}

\cor{The formula for impact ion energy (i.e. $E_0 =  \gamma_i kT_i+3kT_e$ where $\gamma_i = 2$) is good enough for the scope of this paper, and the simulation results match well with experimental data. However, it should be noted that the formula does not include all physical processes for computing the impact ion energy.  Due to the pre-sheath acceleration, the ions at the sheath edge do not have maxwellian velocity distribution. This results in different $\gamma_i$ depending on the model assumed ($\gamma_i \approx 1.5 \sim 2.93$)  \cite{Bissell}.  For more accurate calculation of $\gamma_i$ it would be needed to solve full kinetic equations using the distorted energy distribution of the ions. However, the calculation has not been well established yet. The pre-sheath energy gain is ignored in the adopted formula since it is assumed to be small compared to the uncertainty of $\gamma_i$ and the sheath energy gain. However, the wall impact energy of impurities can be higher than the value calculated using the adopted formula if impurity flow velocity is equal to the ion sound speed $C_s$ ($=\sqrt{\frac{T_e+T_i}{m_D}}$). For example, in the case of $Be^{1+}$ in isothermal plasma ($T_e=T_i$), the wall impact energy is
\begin{eqnarray}
E^{Be^{1+}}_0=0.5 kT_e + \frac{1}{2}m_{Be}C_s^2+(2kT_i+3kT_e) \approx 10 kT_e
\end{eqnarray}
where the first term is due to pre-sheath acceleration, the second term is due to the sound speed equilibrium, and the third term is due to temperature equilibration plus sheath-acceleration. The result is almost twice higher than the formula ($E_0 =  2 kT_i+3kT_e$). Fortunately, the formula used in the simulations is still valid since the majority of impacting ions are deuterium in the simulation.}


The deuterium recycling coefficient has a significant influence on plasma burn-through and the dynamics of deuterium recycling coefficient is different for the carbon wall and the ITER-like wall. This probably results from the different retention of deuterium at the wall. In the carbon wall, deuterium is retained especially after disruptions, so that the simulation results using decay model of $Y_D^D$ match well with JET data \cite{DYON}. However, in the ITER-like wall, the DYON results using the growing model of $Y_D^D$ shows good agreement with JET data. This might be due to the fact that most of attached deuterium, even after disruption event, are easily removed by pumping between discharges \cite{Peter_2012disruption} (typically 30 $\sim$ 40 minutes in JET).

Deuterium fuelling has a significant effect, and it should be confirmed for ITER simulations. However, it might not be a critical issue for plasma burn-through if the fuelling is pre-programmed after about $100[msec]$, when plasma burn-through is completed. As shown in Figure \ref{Scanning_fuelling_82003a}, fuelling does not result in differences on plasma parameters during the plasma burn-through phase. 

As shown in Figure \ref{Scanning_initialC_82003a}(b), the radiation barrier in the ITER-like wall simulations is dominated by deuterium, carbon, and oxygen. The simulation results with \textcolor{correction}{$0.5\%$} $n_C^0(0)$ show good agreement with experimental data. However, the simulation results with  \textcolor{correction}{$0 \%$ and $1\%$} $n_C^0(0)$ still show reasonable agreement in the other plasma paramters, i.e. $I_p(t)$, $T_e(t)$, and $n_e(t)$. 






\cor{Plasma initiation in JET experiments with the ITER-like wall is very reliable. According to an experimental characterisation of plasma formation with the ITER-like wall \cite{Peter_2012IAEA}, the failures during the plasma burn-through phase, that usually occurred with the carbon wall (mostly after disruptions), were not observed with the ITER-like wall. This implies that the plasma parameters during the burn-through phase, including the radiated power losses from impurities, are not likely to vary significantly in each shot, even after disruption events. Hence, the validation of $\# 82003$ is representative of plasma burn-through simulations for other shots with the ITER-like wall.}


\textcolor{correction}{In ITER, oxidation of beryllium surface might be much less than in JET, due to the  much longer plasma pulses. Also, the initial carbon content would be much reduced if the first divertor is made solely of tungsten. On the other hand, due to the use of seeding gases, there should be various impurities e.g. Ar, Ne, or N.  Hence, \cor{modelling impurity seeding in the simulations would be interesting} for predictive simulations of ITER. In DYON simulations, RF heating is not included as no RF heating was used for the simulated pulse ($\#$ 82003), but RF-assisted start-up is planned in ITER. In order to apply the DYON code to such operations, it is required for development of an RF-heating module. Modification of the points stated above will enable the DYON code to be applied for a predictive simulation of the plasma formation in ITER.}

\section{Conclusion}\label{sectionConclusion}

In this article, key physics aspects in the plasma burn-through phase are investigated with the DYON code. The criterion for plasma burn-through is explained with the Radiation and Ionization Barrier (RIB) and the critical degree of ionization $\gamma_{iz}(t_{RIB})$. 

For ITER-like wall simulations, the modification of the \textcolor{correction}{wall-sputtering} model in the DYON code is described in detail, and the simulation results using the \textcolor{correction}{modified wall-sputtering} models are compared to JET data with the ITER-like wall, and the simulation results show good agreement. 

The results of parameter scanning in DYON simulations show that the deuterium recycling coefficient significantly influences the gross energy and particle balance, and the simulation results with the growing model of $Y^D_D$ show good agreement. This implies that during the plasma burn-through phase deuteriums are pumped out by the beryllium wall. Second, \textcolor{correction}{the physical sputtering model using the BeO layer erosion model agrees well with the photomultiplier tube data.} Third, the radiation barrier in the ITER-like wall is dominated by deuterium and other impurities rather than beryllium. The initial carbon content does not influence on other plasma parameters significantly in the simulation results. Fourth, in the case of JET with the ITER-like wall, plasma burn-through is not affected by gas fuelling if the fuelling is pre-programed after 100 $[msec]$.  

The required electric field for deuterium burn-through is calculated by the DYON code, and it is compared to the Townsend avalanche criterion. The limitations set by the burn-through criterion will reduce the operational space with respect to those only based on the Townsend criterion for an electron avalanche. 

The operation space available for JET is computed for the carbon wall and the ITER-like wall. The impurity effects result in a reduced operation space for the carbon wall compared to a pure deuterium plasma. However, the RIB in the ITER-like wall is not much higher than in a pure deuterium plasma. This results in larger operation space available for successful plasma initiation in the ITER-like wall than in the carbon wall. \cor{It is also observed that the prefill gas pressure and toroidal electric field used in successful plasma initiation of JET experiments with the ITER-like wall are located within the operation space computed for the ITER-like wall.}





\section*{Acknowledgement}
This research was funded partly by the Kwanjeong Educational Foundation and by the European Communities under the contract of Association between EURATOM and CCFE. The views and opinions expressed herein do not necessarily reflect those of the European Commission. This work was carried out within the framework of the European Fusion Development Agreement.

\section*{References}

\bibliographystyle{unsrt}

\newpage

\begin{table}[H]
\begin{center}
\begin{tabular}{| c | c |}
\hline
Plasma parameters & Input values \\
\hline
Toroidal magnetic field $B_{\phi}$ & 2.3 $[Tesla]$ \\
\hline
Vertical magnetic field $B_v$ & 0.001 $[Tesla]$ \\
\hline
Initial plasma current density $J_p(0)$ & $ 382.5 \times E = 405.8$ $[A \ m^{-2}] $ \\
\hline
Initial Eddy current  $I_{MK2}(0)$ &  0 $[A] $ \\
\hline
Initial electron temperure $T_e(0)$ & 1 $[eV]$ \\
\hline
Initial ion temperature $T_i(0)$ & 0.03 $[eV]$\\
\hline
Prefilled gas pressure $p(0)$ & Figures \ref{Figure1} and \ref{Figure2} : $5 $ and $7 \ [\times 10^{-5} \ Torr]$   \\  &  Figure \ref{Figure4} : $1$, $3 $, $5 $, and $7$ $ [ \times 10^{-5} \ Torr]$ \\ & Figure \ref{Figure5} :  $5 \times 10^{-5}[Torr]$ \\ & Figure \ref{Figure6} : $1 \times 10^{-6} \sim 2 \times 10^{-4}[Torr]$ \\
\hline
Initial D atom density $n^0_D(0)$ & $2.78 \times 10^{22} \times p(0) [Torr] $ \\
\hline
Initial degree of ionization $\gamma_{iz}(0)$ & $0.002$\\
\hline
Initial impurity $n_I^0(0)$ & $0 $\\
\hline
$Y^D_D$ & $1$\\
\hline
Fuelling efficiency & $0\%$ i.e. No additional fuelling \\
\hline
PSI model & Figure \ref{Figure1}, \ref{Figure2},  and \ref{Figure4} : No PSI effects   \\ & Figure \ref{Figure5} and \ref{Figure6} : No PSI, C wall, or Be wall  \\
\hline
Plasma major radius $R$ & $3 [m]$\\
\hline
Plasma minor radius $a$ & Figure \ref{Figure1}, \ref{Figure2}, and \ref{Figure4} : 0.5 $[m]$ \\ & Figure \ref{Figure5} and \ref{Figure6} :  0.9 $[m]$\\
\hline
Internal inductance $l_i$ & $0.5$\\
\hline
Loop voltage $U_l$ & 20 $[V]$ \\
\hline 
Effective vessel volume & 100 $[m^3]$ \\
\hline
\end{tabular}
\caption{Plasma parameters assumed for DYON simulations. (Figure \ref{Figure1}, \ref{Figure2},  \ref{Figure4}, \ref{Figure5}, and \ref{Figure6}) }  \label{defaultvalues}
\end{center}
\end{table}

\begin{table}[H]
\begin{center}
\begin{tabular}{| c | c | c | c | c | c | c |}
\hline
 Incident ion / target & $D^{1+}$ / Be & $D^{1+}$ / BeO & $Be^{z+}$ / Be & $C^{z+}$/Be & $O^{z+}$ / Be  \\
\hline
$E_{th} [eV]$  & $10$  & $29$  &  $23$ & $40$ & $70$\\
\hline
$E_{TF} [eV]$ & $282$  & $444$  &  $2208$ & $4152$ & $6970$\\
\hline
$Q$ & $0.22$  & $0.13$  &  $0.77$ & $1.6$ & $1.3$\\
\hline
\end{tabular}
\caption{Parameters for physical sputtering yield of beryllium in the ITER-Like Wall \cite{stangeby},\cite{Roth1997}}  \label{BeSputteringParameters}
\end{center}
\end{table}


\begin{table}[H]
\begin{center}
\begin{tabular}{| c | c |}
\hline
Plasma parameters & Input values \\
\hline
Toroidal magnetic field $B_{\phi}$ & 2.7 $[Tesla]$ \\
\hline
Vertical magnetic field $B_v$ & 0.001 $[Tesla]$ \\
\hline
Initial plasma current $I_p(0)$ & 0 $[A]$ \\
\hline
Initial Eddy current  $I_{MK2}(0)$ & 0 $[A] $ \\
\hline
Initial electron temperure $T_e(0)$ & 1 $[eV]$ \\
\hline
Initial ion temperature $T_i(0)$ & 0.03 $[eV]$\\
\hline
Prefilled gas pressure $p(0)$ &   $4.3135 \times 10^{-5}$ $[Torr]$ \\
\hline
Initial Deuterium atom density $n^0_D(0)$ & $2.78 \times 10^{22} \times p(0) [Torr] $ \\
\hline
Initial degree of ionization $\gamma_{iz}(0)$ & $0.002$\\
\hline
Initial Be content $n^0_{Be}(0)$ & $0.01 \times n^0_D(0)$ $[m^{-3}]$\\
\hline
Initial C content $n^0_{C}(0)$ & $0.005 \times n^0_D(0)$ $[m^{-3}]$ \\
\hline
Initial O content $n^0_{O}(0)$ & 0 $[m^{-3}]$  \\
\hline
$Y^D_D$ & $c_1 = 0.9, c_2 = -0.1, c_3= 0.1$ in \Eref{Drecmodel} \\
\hline
Fuelling efficiency & $10$ $\%$ \\
\hline
PSI model & physical sputtering with BeO wall \\
\hline
Plasma major radius $R(t)$ & EFIT $(R(0)=3.0381$ $[m])$\\
\hline
Plasma minor radius $a(t)$ & EFIT $(a(0)=0.08519$ $[m])$\\
\hline
Internal inductance $l_i$ & $0.5$\\
\hline
Loop voltage $U_l(t)$ & Measured in JET \\
\hline 
Vacuum vessel volume & 100 $[m^3]$ \\
\hline
\end{tabular}
\caption{Plasma parameters assumed for the DYON simulation ($\# 82003$) in the ITER-Like Wall}  \label{conditionforILWsim}
\end{center}
\end{table}

\end{document}